\documentclass[showpacs,twocolumn,aps,preprintnumbers,letterpaper]{revtex4}
\usepackage{amsmath,amssymb}
\usepackage{epsfig}
\usepackage{graphicx}
\usepackage{amsmath}
\usepackage{slashed}
\usepackage{amsfonts}
\usepackage{epstopdf}
\newcommand{\be}{\begin{equation}}
\newcommand{\ee}{\end{equation}}
\newcommand{\bear}{\begin{eqnarray}}
\newcommand{\eear}{\end{eqnarray}}
\newcommand{\ba}{\begin{array}}
\newcommand{\ea}{\end{array}}

\def\be{\begin{eqnarray}}
\def\ee{\end{eqnarray}}
\def\bea{\be}
\def\eea{\ee}

\def\roughly#1{\mathrel{\raise.3ex\hbox{$#1$\kern-.75em%
\lower1ex\hbox{$\sim$}}}}

\begin{document}

\title{Heavy Baryons and their Exotics from Instantons in Holographic QCD}

\author{Yizhuang Liu}
\email{yizhuang.liu@stonybrook.edu}
\affiliation{Department of Physics and Astronomy, Stony Brook University, Stony Brook, New York 11794-3800, USA}

\author{Ismail Zahed}
\email{ismail.zahed@stonybrook.edu}
\affiliation{Department of Physics and Astronomy, Stony Brook University, Stony Brook, New York 11794-3800, USA}

%%%%%%%%%

\date{\today}
\begin{abstract}
We use a variant of the $D4$-$D8$ construction that includes two chiral and one heavy meson,
to describe heavy-light baryons and their exotics as heavy mesons bound to a flavor instanton in bulk. 
At strong coupling, the heavy meson is shown to {\it always bind} in the form of a flavor instanton
zero mode in the fundamental
representation. The ensuing instanton moduli for the heavy baryons exhibits both chiral and 
heavy quark symmetry.  We detail how to quantize it, and
derive model independent mass relations for heavy bayons
with a single-heavy quark in leading order, in overall agreement with the reported baryonic spectra with one
charm or bottom. We also discuss the low-lying masses and quantum assignments for the even and odd 
parity states, some of which are yet to be observed. We extend our
analysis to double-heavy pentaquarks with hidden charm and bottom. In leading order, we find a pair of doube-heavy iso-doublets 
with $IJ^\pi=\frac 12 {\frac 12}^-, \frac 12 {\frac 32}^-$ assignments for all heavy flavor combinations. 
We also predict five new Delta-like pentaquark states with 
$IJ^\pi=\frac 32 {\frac 12}^-, \frac 32 {\frac 32}^-, \frac 32 {\frac 52}^-$ assignments for both charm and bottom.
 \end{abstract}
%\pacs{11.25.Tq, 13.60.Hb,13.85.Lg}
%11.25.Tq 	Gauge/string duality  13.60.Hb for deep-inelastic structure functions; 13.85.Lg 	Total cross sections

%11.15.Kc	Classical and semiclassical techniques
%11.30.Rd	Chiral symmetries
%12.38.Lg	Other nonperturbative calculations

\maketitle

\setcounter{footnote}{0}

%\baselineskip 18pt \pagebreak
%\renewcommand{\thepage}{\arabic{page}}
%\tableofcontents
%\pagebreak

\section{Introduction}
In QCD the light quark sector (u, d, s) is dominated by the spontaneous breaking of chiral
symmetry. The heavy quark sector (c, b, t) is characterized by heavy-quark symmetry~\cite{ISGUR}. 
The combination of both symmetries is at the origin of the chiral doubling in heavy-light mesons~\cite{MACIEK,BARDEEN}
as measured by both the BaBar collaboration~\cite{BABAR}  and the CLEOII collaboration~\cite{CLEOII}.

Recently the Belle collaboration~\cite{BELLE} and the BESIII collaboration~\cite{BESIII} have reported many
multiquark exotics uncommensurate with quarkonia, e.g.  the neutral $X(3872)$ and the charged $Z_c(3900)^\pm$ 
and $Z_b(10610)^\pm$.  These exotics have been also confirmed 
 by the DO collaboration at Fermilab~\cite{DO},  and the  LHCb collaboration at CERN~\cite{LHCb}. 
LHCb has reported new pentaquark states $P_c^+(4380)$ and $P_c^+(4450)$ through the
decays $\Lambda_b^0\rightarrow J\Psi pK^-, J\Psi p\pi^-$~\cite{LHCbx}. More recently, five narrow and neutral excited $\Omega^0_c$
baryon states that decay primarily to  $\Theta_c^+K^-$ were also reported by the same collaboration~\cite{LHCbxx}.
These flurry of experimental results  support  new phenomena involving heavy-light multiquark states, a priori  outside the 
canonical classification of the quark model.

Some of the tetra-states exotics maybe understood as
molecular bound states mediated by one-pion exchange much
like deuterons or deusons~\cite{MOLECULES,THORSSON,KARLINER,OTHERS,OTHERSX,OTHERSZ,OTHERSXX,LIUMOLECULE}.
Non-molecular heavy exotics were also discussed using constituent quark models~\cite{MANOHAR}, 
heavy solitonic baryons~\cite{RISKA,MACIEK2}, instantons~\cite{MACIEK3} and QCD sum rules~\cite{SUHONG}. 
The penta-states exotics reported in~\cite{LHCbx} have been foreseen in~\cite{MAREK} and since addressed by many
using both molecular and diquark constructions~\cite{MANY},  as well as a bound anti-charm to a Skyrmion~\cite{PENTARHO}.
String based pictures using string junctions~\cite{VENEZIANO} have also been suggested for the description of exotics,
including a recent proposal  in the context of the holographic inspired string hadron model~\cite{COBI}.

The holographic construction offers a framework for addressing both chiral symmetry and confinement
in the  double limit of large $N_c$ and large t$^\prime$Hooft coupling $\lambda=g^2N_c$. A concrete 
model was proposed by Sakai and Sugimoto~\cite{SSX}  using a $D4$-$D8$ brane construction. The
induced gravity on the probe $N_f$ $D8$ branes due to the large stack of $N_c$ $D4$ branes, causes
the probe branes to fuse in the holographic direction, providing a geometrical mechanism for the spontaneous
breaking of chiral symmetry.  The DBI action on the probe branes yields a low-energy effective
action for the light pseudoscalars with full global chiral symmetry, where the vectors and axial-vector light mesons 
are dynamical gauge particles of a hidden chiral symmetry~\cite{HIDDEN}. In the model, light
baryons are identified  with small size instantons by wrapping $D4$ around $S^4$,  and are dual to Skyrmions on 
the boundary~\cite{SSXB,SSXBB}. Remarkably, this identification provides a geometrical description of the baryonic core that
is so elusive in most Skyrme models~\cite{SKYRME}. A first principle description of the baryonic core is paramount to the understanding
of heavy hadrons and their exotics since the heavy quarks bind over their small Compton wavelength.

The purpose of this paper is to propose a holographic description of heavy baryons and their exotics that involve
light and heavy degrees of freedom through a variant of the $D4$-$D8$ model  that includes
a heavy flavor~\cite{LIUHEAVY} with both chiral and heavy-quark symmetry. The model uses 2 light and 1 heavy branes where
the heavy-light mesons are identified with the string low energy modes, and approximated
by bi-fundamental and local  vector fields in the vicinity of the light probe branes. Their masses follow from the vev of the
moduli span by the dilaton fields in the DBI action. The model allows for the description of the radial spectra of the
$(0^\pm, 1^\pm)$ heavy-light multiplets, their pertinent vector and axial correlations, and leads reasonable estimates
for the one-pion axial couplings and radiative decays in the heavy-light sector.

In this construction, the heavy baryons will be sought in the form of a
bulk instanton in the worldvolume of $D8$ bound to heavy-light vector mesons,
primarily the heavy-light $(0^-,1^-)$ multiplet. This approach
will extend the bound state approach developed in the context of the Skyrme model~\cite{PENTARHO,SKYRMEHEAVY} 
to holography. We note that alternative holographic models for the description of heavy hadrons have been
developed  in~\cite{FEWX,BRODSKY} without the dual strictures of chiral and heavy quark symmetrty.

The organization of the paper is as follows: In section 2 we 
briefly outline  the geometrical set up  for the derivation of the heavy-light  effective action  
through the pertinent bulk DBI and CS actions. 
In section 3,  we detail the heavy-meson interactions to the flavor instanton in bulk. 
In section 4,  we show how a vector meson with  spin 1 binding to the bulk instanton  transmutes
to a spin $\frac 12$. In section 5,  we identify the moduli  of the bound zero-mode and quantize it by collectivizing
some of the soft modes. The mass spectra for baryons with single- and double-heavy quarks are explicitly
derived. Some of our exotics are comparable to those recently reported by several collaborations, while others are new.
Our conclusions are in section 6. In the Appendix we briefly review the quantization of the light meson moduli without
the heavy mesons.

\section{ Holographic effective action}

\subsection{D-brane set up}

\begin{figure}[h!]
\begin{center}
 \includegraphics[width=6cm]{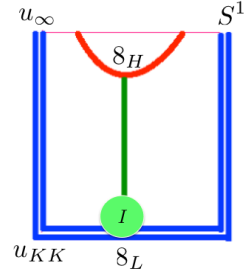}
  \caption{$N_f-1=2$ antipodal $8_L$ light branes, and one $8_H$  heavy brane shown in the $\tau U$ plane,
  with a bulk $SU(2)$ instanton embedded in $8_L$ and a massive $HL$-string connecting them.}
 \label{fig_branex}
 \end{center}
\end{figure}

The $D4$-$D8$ construction proposed by Sakai and Sugimoto~\cite{SSX}
for the description of the light hadrons  is standard and will not be repeated here. Instead, we follow~\cite{LIUHEAVY} and
consider the variant with  $N_f-1$ light $D8$-$\bar D8$ (L) and one heavy (H) probe branes in the cigar-shaped geometry that
spontaneously breaks chiral symmetry. For simplicity, the light probe branes are always assumed
in the anti-podal configuration. A schematic description  of the set up for $N_f=3$ is shown in Fig.~\ref{fig_branex}.
We assume that the L-brane world volume consists of $R^{4}\times S^1\times S^4$ with 
$[0-9]$-dimensions.  The light 8-branes are embedded in the $[0-3+5-9]$-dimensions and set
at the antipodes of $S^1$ which lies in the 4-dimension. 
The warped $[5-9]$-space is characterized by a finite size  $R$ and a horizon at $U_{KK}$.

\subsection{DBI and CS actions}

The lowest open string modes streched between the H- and L-branes  are attached to a wrapped $S^4$ in $D4$ 
shown as an instanton in~Fig.~\ref{fig_branex}. Near the L brane world volume, these string modes 
consist of tranverse modes $\Phi_M$ and longitudinal modes $\Psi$, both fundamental with respect to the flavor group $SU(N_f-1)$. 
At non-zero brane separation, these fields acquire  a vev that makes the vector field massive~\cite{MEYERS}.  Strictly speaking these
fields are bi-local, but near the L-branes we will approximate them  by local vector fields that are described by the standard DBI action
in the background of a warped instanton field. 
In this respect, our construction is distinct from the approaches developed  in~\cite{FEWX}.

With this in mind and to leading order in the $1/\lambda$ expansion, the effective action on the probe L-branes
consists of the non-Abelian DBI  (D-brane Born-Infeld) and CS (Chern-Simons) action.  After integrating over the $S^4$, the leading contribution to the DBI action is

\bea
\label{1}
S_{\rm DBI}\approx -\kappa\int d^4x dz\,{\rm Tr}\left({\bf f}(z){\bf F}_{\mu\nu}{\bf F}^{\mu\nu}+{\bf g}(z){\bf F}_{\mu z}{\bf F}^{\nu z}\right)
\eea
Our conventions are $(-1,1,1,1)$ with $A_{M}^{\dagger}=-A_M$. The warping factors are 

\be
{\bf f}(z)=\frac{R^3}{4U_z}\,,\qquad {\bf g}(z)=\frac{9}{8}\frac{U_z^3}{U_{KK}}
\ee 
with $U_z^3=U_{KK}^3+U_{KK}z^2$, $\kappa=\tilde T(2\pi \alpha^\prime)=a\lambda N_c$ and $a=1/(216\pi^3)$~\cite{SSX}.
All dimensions are understood in units where the Kaluza-Klein mass $M_{KK}\equiv 1$ unless specified otherwise.
The effective fields in the field strengths are
($M,N$ run over $(\mu,z)$)

\bea
\label{2}
&&{\bf F}_{MN}=\nonumber \\ 
&&\left(\begin{array}{cc}
F_{MN}-\Phi_{[M}\Phi_{N]}^{\dagger}&\partial_{[M}\Phi_{N]}+A_{[M}\Phi_{N]}\\
-\partial_{[M}\Phi^{\dagger}_{N]}-\Phi^{\dagger}_{[M}A_{N]}&-\Phi^{\dagger}_{[M}\Phi_{N]}
\end{array}\right)
\eea
The CS contribution to the effective action is (form notation used)

\be
\label{6}
S_{\rm CS}=\frac{N_c}{24\pi^2}\int_{R^{4+1}}{\rm Tr}\left({\bf A}{\bf F}^2-\frac{1}{2}{\bf A}^3{\bf F}+\frac{1}{10}{\bf A}^5\right)
\ee
where the normalization to $N_c$ is fixed by integrating the $F_4$ RR flux over the $S^4$.
The  matrix valued 1-form gauge field is
\be
\label{7}
{\bf A}=\left(\begin{array}{cc}
A&\Phi\\
-\Phi^{\dagger}&0
\end{array}\right)
\ee

For $N_f$ coincidental branes, the $\Phi$ multiplet is massless. However, their brane world-volume 
supports an adjoint and traceless scalar $\Psi$ in addition to the adjoint gauge field $A_M$ both of which are
hermitean and $N_f\times N_f$ valued, which we have omitted from the DBI action  in so far
for simplicity. They are characterized by a quartic potential with finite extrema and a vev $v$ for 
the diagonal of $\Psi$~\cite{MEYERS}.
As a result  the $\Phi$ multiplet acquires a Higgs-like mass of the type

\bea
\label{8X3}
\frac 12 m_H^2 {\rm Tr}\left(\Phi^\dagger_M \Phi_M\right)\sim \frac 12 v^2{\rm Tr}\left(\Phi^\dagger_M \Phi_M\right)
\eea
The vev is related to the separation between the light and heavy branes~\cite{MEYERS}, 
which we take it to be the mass following from the length of the streched HL string, and which 
we identify as the mass of the heavy-light $(0^-, 1^-)$ multiplet  for either charm
$(D,D^*)$ or bottom $(B,B^*)$. 
 In the heavy quark limit, the radial spectra, axial and vector correlations, and 
the one-pion radiative decays of the $(0^-,1^-)$ multiplet  are fairly reproduced by this model~\cite{LIUHEAVY}. 
%Below, $m_Q$ will be identified 
%mass for the bi-fundamental field $\Phi$ which will be identified with the HL meson doublets,
%such as   $(D, D^*)$ or  $(B, B^*)$. Throughout, we will refer to the large mass limit also as 
%the heavy quark limit. 

%%%%%%%%%%%%%%%%%%%%
\section{Heavy-Light-Instanton  interactions}

In the original two-flavor $D4$-$D8$ set up by Sakai and Sugimoto~\cite{SSX} light baryons  are  first identified with a flavor
instanton in bulk~\cite{SSXB} and its moduli  quantized to yield the nucleon and Delta~\cite{SSXBB}. This construction holds
in our case in the light sector of (\ref{1}) verbatum and we refer the interested reader to~\cite{SSXB,SSXBB} for the details
of the analysis. The key observations is that the instanton size is small at strong coupling  $\rho\sim1/\sqrt{\lambda}$, as a 
result of balancing the large and leading attraction due to gravity in bulk (large warpings) and the subleading 
U(1) Coulomb-like repulsion induced by the Chern-Simons term.

In the geometrical set up described in Fig.~\ref{fig_branex}, 
the small size  instanton translates to a flat space 4-dimensional  instanton~\cite{SSXB} 

\begin{eqnarray}
\label{INST}
A^{cl}_{M}=&&-\bar\sigma_{MN}\frac{x_N}{x^2+\rho^2},\nonumber\\
%F_{MN}=&&2i\frac{\bar\sigma_{MN} \rho^2}{(x^2+\rho^2)^2}\nonumber\\
A^{cl}_0=&&\frac{-i}{8\pi^2a x^2}\left(1-\frac{\rho^4}{(x^2+\rho^2)^2}\right)
\end{eqnarray}
after using the rescalings 

\begin{eqnarray}
\label{RESCALE}
&&x_0\rightarrow x_0, x_{M}\rightarrow x_{M}/\sqrt{\lambda}, \sqrt{\lambda}\rho\rightarrow \rho\nonumber\\
&&(A_{0},\Phi_{0})\rightarrow (A_0,\Phi_0), \nonumber\\
&&(A_{M},\Phi_M)\rightarrow \sqrt{\lambda}(A_M,\Phi_M)
\end{eqnarray}
in (\ref{1}).  From here and throughout the rest of the paper, $M,N$ run only over $1,2,3,z$.
%To analyze the interactions of the light and heavy mesons around the small size
%instanton (\ref{INST}), it is useful to perform the following rescalings
To order $\lambda^0$ the rescaled contributions describing the interactions between the light gauge fields $A_M$ and
the heavy fields $\Phi_M$  to quadratic order split in the form 

\be
\label{RS1}
S=aN_c\lambda S_{0}+aN_cS_{1}+S_{CS}
\ee
with each contribution given by

\begin{eqnarray}
\label{RX0}
S_0=&&\nonumber -(D_{M}\Phi_{N}^{\dagger}-D_{N}\Phi_{M}^{\dagger})(D_M\Phi_N-D_N\Phi_M)\nonumber \\  
&&+2\Phi_{M}^{\dagger}F_{MN}\Phi_{N}\nonumber\\
S_1=&&+2(D_{0}\Phi_{M}^{\dagger}-D_{M}\Phi_{0}^{\dagger})(D_0\Phi_M-D_M\Phi_0)\nonumber \\ 
&&-2\Phi_0^{\dagger}F^{0M}\Phi_{M}-2\Phi_{M}^{\dagger}F^{M0}\Phi_0\nonumber\\
&&-2m_H^2\Phi_{M}^{\dagger}\Phi_M +\tilde S_1\nonumber \\
S_{CS}=&&-\frac{iN_c}{24\pi^2}(d\Phi^{\dagger}Ad\Phi+d\Phi^{\dagger}dA\Phi+\Phi^{\dagger}dAd\Phi)\nonumber \\
&&-\frac{iN_c}{16\pi^2}(d\Phi^{\dagger} A^2\Phi+\Phi^{\dagger}A^2d\Phi+\Phi^{\dagger}(AdA+dAA)\Phi)\nonumber \\
&&-\frac{5iN_c}{48\pi^2}\Phi^{\dagger}A^3\Phi+S_C(\Phi^4,A)
\end{eqnarray}
 and 

\begin{eqnarray}
\tilde S_1=&&+\frac{1}{3}z^2(D_i\Phi_j-D_j\Phi_i)^{\dagger}(D_i\Phi_j-D_j\Phi_i)\nonumber \\
&&-2z^2(D_i\Phi_z-D_z\Phi_i)^{\dagger}(D_i\Phi_z-D_z\Phi_i)\nonumber\\
&&-\frac{2}{3}z^2\Phi_i^{\dagger}F_{ij}\Phi_j+2z^2(\Phi_z^{\dagger}F_{zi}\Phi_i+{\rm c.c.})
\end{eqnarray}

\section{Bound State as a Zero-Mode}

We now show that in the double limit of large $\lambda$ followed by large $m_Q$, a heavy meson in bulk
always binds to the flavor instanton  in the form of a 4-dimensional ($123z$) flavor zero-mode that effectively is
a spinor.  This holographic zero-mode translates equally to either a bound heavy flavor or anti-heavy flavor in 
our space-time ($0123$). This is remarkable to holography, as the heavy bound states in the Skyrme-type involve
particles but with difficulties anti-particles~\cite{SKYRMEHEAVY,THETA}. Indeed, in the 
 Skyrme model, the Wess-Zumino-Witten
term which is time-odd, carries opposite signs for heavy particles and anti-particles that are magnified by $N_c$ 
in comparison to the heavy-mesonic action.  As a result  the particle state is attractive, while the anti-particle state is 
 repulsive.

\subsection{Field equations}

We now consider the bound state solution of the heavy meson field $\Phi_M$ in the (rescaled) instanton background
{\ref{INST}). We note that  the field equation for $\Phi_M$ is independent of $\Phi_0$ and reads

\be
\label{RX2}
D_MD_M\Phi_N+2F_{NM}\Phi_M-D_{N}D_M\Phi_M=0
\ee
while the contraint field equation (Gauss law) for $\Phi_0$  depends on  $\Phi_M$ through the Chern-Simons term 

\begin{eqnarray}
\label{RX3}
&&D_M(D_0\Phi_M-D_M\Phi_0)\nonumber\\&&-F^{0M}\Phi_M-\frac{\epsilon_{MNPQ}}{64\pi^2a}K_{MNPQ}=0
\end{eqnarray}
with $K_{MNPQ}$ defined as 

 \begin{eqnarray}
 \label{RX4}
 K_{MNPQ}=&&+\partial_{M}A_N\partial_P\Phi_Q+A_MA_N\partial_P\Phi_Q\nonumber\\
 &&+\partial_MA_NA_P\Phi_Q+\frac{5}{6}A_{M}A_NA_P\Phi_Q
 \end{eqnarray}
In the heavy quark limit it is best to redefine 
$\Phi_{M}=\phi_{M}e^{-im_Hx_0}$ for particles. The anti-particle case follows through $m_Q\rightarrow -m_H$
with pertinent sign changes. As a result, the preceding field equations remain unchanged for $\phi_M$ with the
substitution $D_0\phi_M\rightarrow (D_0\mp im_H)\phi_M$ understood for particles $(-)$ or anti-particles ($+$) respectively.

\subsection{Double limit}

In the double limit of $\lambda\rightarrow \infty$ followed by $m_H\rightarrow \infty$, the leading contributions
are of order $\lambda m_H^0$ from the light effective action in (\ref{1}), and of order $\lambda^0m_H$ 
from the heavy-light interaction term $S_1$ in (\ref{RX0}). This double limit is justified if we note that in leading order,
the mass of the heavy meson  follows from the straight pending string shown in Fig~\ref{fig_branex}, with a value~\cite{LIUHEAVY} 

\be
\label{00}
\frac{m_H}{\lambda M_{KK}}= \frac{2}{9\pi } (M_{KK}u_H)^{\frac{2}{3}}
\ee
where $u_H$ is the holographic
height of the heavy brane. The  double limit requires the ratio in (\ref{00}) to be
parametrically small. 

With the above in mind, we have

\be
\label{RX5}
\frac{S_{1,m}}{aN_c}=4im_H\phi^{\dagger}_{m}D_0\phi_{m}-2im_H(\phi_{0}^{\dagger}D_{M}\phi_{M}-{\rm c.c.})
\ee 
and from the Chern-Simons term in (\ref{RX0}) we have

\be
\label{RX6}
\frac{m_H N_c}{16\pi^2}\epsilon_{MNPQ}\phi^{\dagger}_{M}F_{NP}\phi_{Q}=\frac{m_HN_c}{8\pi^2}\phi^{\dagger}_{M}F_{MN}\phi_{N}
\ee
The constraint equation (\ref{RX3}) simplifies considerably to order $m_Q$,

\be
\label{RX7}
D_{M}\phi_{M}=0
\ee
implying that $\phi_M$ is covariantly transverse in leading order in the double limit.

\subsection{Vector to spinor zero-mode}

%Throughout this section $M,N$ will be understood as $\mu,\nu=1,2,3,z$.
The instanton solution $A_M$ in (\ref{INST}) carries a field strength

\begin{eqnarray}
\label{INSTX}
%A_{M}=&&-\bar\sigma_{MN}\frac{x_N}{x^2+\rho^2},\nonumber\\
F_{MN}=&&\frac{2\,\bar\sigma_{MN} \rho^2}{(x^2+\rho^2)^2}
%A_0=&&\frac{1}{8\pi^2a x^2}\left(1-\frac{\rho^2}{(x^2+\rho^2)^2}\right)
\end{eqnarray}
We now observe that the heavy field equation (\ref{RX2}) in combination with the constraint
equation (\ref{RX7}) are equivalent to the vector zero-mode equation in the fundamental representation.
To show that, we recall that the field strength (\ref{INSTX}) is self-dual, and $S_0$ in (\ref{RX0}) can be
written in the compact form

\begin{eqnarray}
\label{RX8}
S_0=&&-f_{MN}^{\dagger}f_{MN}+2\phi_{M}^{\dagger}F_{MN}\phi_{N}\nonumber \\=&&-f_{MN}^{\dagger}f_{MN}+2\epsilon_{MNPQ}\phi_{M}^{\dagger}D_{M}D_{Q}\phi_{N}\nonumber \\=&&-f_{MN}^{\dagger}f_{MN}+f_{MN}^{\dagger}\star f_{MN}\nonumber \\
=&&-\frac{1}{2}(f_{MN}-\star f_{MN})^{\dagger}(f_{MN}-\star f_{MN})
\end{eqnarray}
after using the Hodge dual $\star$ notation,  and defining

\label{RX88}
\be
f_{MN}=\partial_{[M}\phi_{N]}+A_{[M}\phi_{N]}
\ee
Therefore, the second order field equation (\ref{RX2})  can be replaced by the anti-self-dual condition (first order)
and the transversality condition (\ref{RX7}) (first order),

\begin{eqnarray}
\label{RX9}
f_{MN}-\star f_{MN}=0\nonumber\\
D_{M}\phi_{M}=0
\end{eqnarray}
which are equivalent to

\be
\label{RX10}
\sigma_{M}D_{M}\psi= D \psi =0 \qquad{with}\qquad
\psi=\bar \sigma_{M}\phi_{M}
\ee
The spinor zero-mode $\psi$  is unique, and its explicit matrix form reads

\be
\label{RX11}
\psi^{a}_{\alpha \beta}=\epsilon_{\alpha a}\chi_{\beta}\frac{\rho}{(x^2+\rho^2)^{\frac{3}{2}}}
\ee
which gives the vector zero-mode in the form

\be
\label{RX12}
\phi_{M}^a=\chi_{\beta}(\sigma_{M})_{\beta \alpha}\epsilon _{\alpha a}\frac{\rho}{(x^2+\rho^2)^{\frac{3}{2}}}
\ee
or in equivalent column form 

\be
\label{RX13}
\phi_{M}=\bar \sigma_{M}\chi\frac{\rho}{(x^2+\rho^2)^{\frac{3}{2}}}\equiv \bar \sigma_{M}f(x)\chi
\ee
Here $\chi_{\alpha}$ is a constant two-component spinor. It can be checked explicitly that (\ref{RX13})
is a solution to the first order equations (\ref{RX9}).  The interplay between  (\ref{RX11}) and (\ref{RX12})
is remarkable as it shows that in holography a heavy vector meson binds to an instanton in bulk in the
form of a vector  zero mode that is equally described as a spinor. This  duality 
illustrates the transmutation from a spin 1 to a spin $\frac 12$ in the instanton field.

\section{Quantization}

Part of the classical moduli of the bound instanton-zero-mode breaks rotational and translational symmetry,
which will be quantized by slowly rotating or translating the bound state. In addition, it was noted in~\cite{SSXB}
that while the deformation of the instanton size  and holographic location are not collective per say as they 
incur potentials, they are still soft in comparison to the more massive quantum excitations in bulk and should
be quantized as well. The ensuing quantum states are vibrational and identified with the breathingh modes
(size vibration) and odd parity states (holographic vibration).

\subsection{Collectivization}

The leading $\lambda N_c$ contribution is purely instantonic and
its quantization is standard and can be found in~\cite{SSXBB}.
For completeness we have summarized it in the Appendix. The quantization of the
subleading $\lambda^0 m_H$ contribution involves the zero-mode and is new,
so we will describe iin more details.  
For that, we let the zero-mode  slowly translates, rotates and deforms through

\begin{eqnarray}
\label{RX14}
\Phi\rightarrow &&V({ a}_I(t)) \Phi(X_0(t),Z(t),\rho(t),\chi(t))\nonumber\\
\Phi_0\rightarrow &&0+\delta \phi_0
\end{eqnarray}
Here $X_0$ is the center in the 123 directions and Z is the center in the  $z$ directon. $a_I$ is the SU(2) gauge rotation moduli. We denote the moduli   by  $X_{\alpha}\equiv(X,Z,\rho)$ with

\begin{eqnarray}
\label{RX15}
-iV^{\dagger}\partial_{0}V=&&\Phi=-\partial_{t}X_{N}A_{N}+\chi^{a}\Phi_{a}\nonumber\\
\chi^{a}=&&-i{\rm Tr}\left(\tau^a\,{a}_I^{-1}\partial_{t}\,{a}_I\right)
\end{eqnarray}
$a_I$ is the SU(2) rotation which carries the isospin and angular momentum quantum numbers. 
%The $D_{M}D_{M}\Phi^{a}=0$ in singular gauge  are proportional to $\tau_a$. 
The constraint equation
(\ref{RX3})  for $\phi_0$ has to be satisfied, which fixes  $\delta \phi_0$ at sub-leading order

\begin{eqnarray}
\label{RX16}
&&-D_{M}^2\delta \phi_0+D_{M}\bar\sigma_{M}(\partial_{t}X_i\partial_{X_i}f\chi+\partial_{t}\chi)\nonumber \\ 
&&+i(\partial_tX_\alpha \partial_{\alpha}\Phi_{M}-D_M\Phi)\bar\sigma_{M}\chi+\delta S_{cs}=0
\end{eqnarray}
The solution to (\ref{RX16}) can be inserted back into the action for a general
quantization of the ensuing moduli.

\subsection{Leading heavy mass terms}

There are three contributions to order $\lambda^0m_H$,  namely 

\be
\label{RX17}
16im_H\chi^{\dagger}\partial_{t}\chi f^2  + 16im_H \chi^{\dagger}\chi A_0 f^2 -m_Hf^2\chi^{\dagger}\sigma_{\mu}\Phi\bar \sigma_{\mu}\chi
\nonumber\\
\ee
with the rescaled U(1) field $A_0$,  and the Chern-Simons  term

\be
\label{RX18}
%\frac{m N_c}{16\pi^2}\epsilon_{MNPQ}\phi^{\dagger}_{M}F_{NP}\phi_{Q}=
\frac{im_HN_c}{8\pi^2}\phi^{\dagger}_{M}F_{MN}\phi_{N}=\frac{i3m_HN_c}{\pi^2}\frac{f^2\rho^2}{(x^2+1)^2}\chi^{\dagger}\chi
\ee
with the field strength given in (\ref{INSTX}). Explicit calculations show that the third contribution in (\ref{RX17})
vanishes owing to the identity  $\sigma_{\mu}\tau_a\bar \sigma_{\mu}=0$.

The coupling $\chi^\dagger \chi A_0$ term in 
(\ref{RX17}) induces a Coulomb-like back-reaction. To see this, we set $\psi=iA_0$ and collect all
 the U(1) Coulomb-like  couplings in  the rescaled effective action to order $\lambda^0m_H$

\bea
\label{COULOMB}
\frac{S_C(A_0)}{aN_c}=&&\int\left(\frac{1}{2}(\nabla \psi)^2+\psi(\rho_0[A]-16m_Hf^2\chi^{\dagger}\chi)\right)\nonumber\\
\rho_0[A]=&&\frac{1}{64\pi^2a}\epsilon_{MNPQ}F_{MN}F_{PQ}
\eea
The static action contribution stemming from the coupling to the U(1) charges $\rho_0$ and $\chi^\dagger \chi$
is

\be
\frac{S_C}{aN_c}\rightarrow \frac{S_C[\rho_0]}{aN_c}+16m_{H}\chi^\dagger \chi\int f^2(-iA_0^{cl})-\frac{(16m_{H}\chi^{\dagger}\chi)^2}{24\pi^2}\nonumber\\
\ee
The last contribution is the Coulomb-like self-interaction induced by the instanton on the heavy meson through the
U(1) Coulomb-like field in bulk. It is repulsive and tantamount of fermion number repulsion in holography.

\subsection{Moduli effective action}

Putting all the above contributions together, we obtain the effective action density on the moduli in leading order in the
heavy meson mass

\begin{eqnarray}
\label{RX19}
&&{\cal L}={\cal L}_0[a_I, X_\alpha]
+16aN_cm_H\left(i\chi^{\dagger}\partial_0\chi^{\dagger}\int d^4x\, f^2\right.\nonumber\\
&&\left.-\chi^{\dagger}\chi\int d^4x\, f^2\left(iA_0^{cl}-\frac{3}{16a\pi^2}\frac{\rho^2}{(x^2+\rho^2)^2}\right)\right)\nonumber \\&&-aN_c\frac{(16m_{H}\chi^{\dagger}\chi)^2}{24\pi^2\rho^2}
\end{eqnarray}
with ${\cal L}_0$ referring to the effective action density on the moduli stemming from the contribution
of the light degrees of freedom in the instanton background. It is identical to the one
derived in~\cite{SSXB} and to which we refer the reader for further details.  In (\ref{RX19})
We have made explicit the new contribution due to the bound heavy meson through $\chi$. To this order there
is no explicit coupling of the light collective degrees of freedom $a_I,$ to the heavy
spinor degree of freedom $\chi$, a general reflection on heavy quark symmetry in leading order.
However, there is a coupling to the instanton size $\rho$ through the holographic direction which
does not upset this symmetry.
After using the normalization $\int d^4x\,f^2=1$,   inserting the explicit form of  $A^{cl}_0$
from (\ref{INST}), and rescaling $\chi \rightarrow {\chi}/{2\sqrt{aN_cm_H}}$, we finally have

\be
\label{RX20}
{\cal L}={\cal L}_0[a_I, X_\alpha]+\chi^{\dagger}i\partial_t\chi +\frac{3}{32\pi^2a\rho^2}\chi^{\dagger}\chi-\frac{(\chi^{\dagger}\chi)^2}{24\pi^2a \rho^2N_c}\nonumber\\
\ee
Remarkably, the bound vector zero-mode to the instanton transmutes to a massive spinor with a repulsive 
Coulomb-like self- interaction. The mass is {\it negative} which implies that the heavy meson lowers its energy in 
the presence of the instanton to order $\lambda^0$. We note that the preceding arguments carry verbatum 
to an anti-heavy meson in the presence of an instanton, leading (\ref{RX20}) with a {\it positive} mass term. This meson raises
its energy in the presence of the instanton to order $\lambda^0$. These effects originate from the 
Chern-Simons action  in holography. They are the analogue of the effects due to the 
Wess-Zumino-Witten  term in the standard Skyrme model~\cite{SKYRMEHEAVY,THETA}. While they are
leading in $1/N_c$ in the latter causing the anti-heavy meson to unbind in general, they are subleading 
in $1/\lambda$ in the former where to leading order the bound state is always a  BPS zero mode irrespective
of heavy-meson or anti-heavy-meson.

\subsection{Heavy-light spectra}

The quantization of (\ref{RX20}) follows the same arguments as those presented in~\cite{SSXB}
for ${\cal L}_0[a_I, X_\alpha]$ and to which we refer for further details in general, and the Appendix for the notations in particular. 
Let $H_0$ be the Hamiltonian 
associated to ${\cal L}_0[a_I, X_\alpha]$, then the Hamiltonian for (\ref{RX20})  follows readily in the form

\be
\label{RX21}
H=H_0[\pi_I,\pi_X, a_I, X_\alpha]-\frac{3}{32\pi^2a\rho^2}\chi^{\dagger}\chi+\frac{(\chi^{\dagger}\chi)^2}{24\pi^2a \rho^2N_c}\nonumber\\
\ee
with the new quantization rule for the spinor 

\be
\label{RX22}
\chi_i\chi_j^{\dagger}\pm\chi_j^{\dagger}\chi_i=\delta_{ij}
\ee
The statistics of $\chi$ needs to be carefully determined. For that, we note the symmetry transformation

\be
\label{RX23}
\chi\rightarrow  U\chi\qquad and\qquad 
\phi_{M}\rightarrow U \Lambda_{MN}\phi_{N}
\ee
since $U^{-1}\bar \sigma_{M}U=\Lambda_{MN}\bar\sigma_{N}$. So a rotation of the spinor $\chi$ 
is equivalent to a spatial rotation of the heavy vector meson field $\phi_M$ which carries spin 1. Since $\chi$
is in the spin $\frac 12$ representation it should be quantized as a fermion. So only the plus sign is to be
retained in (\ref{RX22}). Also, $\chi$ carries opposite parity  to $\phi_M$, i.e. positive.
With this in mind, the spin ${\bf J}$ and isospin ${\bf I}$ are then related by

\be
\label{RX24}
\vec{\bf J}=-\vec{\bf I}+\vec{\bf S}_\chi\equiv -\vec{\bf I}+ \chi^\dagger\frac{\vec\tau}2\chi
\ee
We note that in the absence of the heavy-light meson ${\bf J}+{\bf I}=0$ as expected from the spin-flavor
hedgehog character of the bulk  instanton (see also the Appendix). 

The spectrum of (\ref{RX21})  follows from the one discussed in details in~\cite{SSXB} with the 
only  modification of $Q$ entering in $H_0$ as given  in the Appendix

\be
\label{RX25}
Q\equiv \frac{N_c}{40a\pi^2}\rightarrow \frac{N_c}{40a\pi^2} \left(1-\frac {15}{4N_c} \chi^{\dagger}\chi +\frac{5(\chi^{\dagger}\chi)^2}{3N_c^2} \right)
\ee
The quantum states with a single bound state $N_Q=\chi^\dagger \chi=1$ and $IJ^\pi$ assignments are labeled by

\be
\label{RX26}
\left| N_Q,JM,lm,n_z,n_\rho\right>\,\,\,with \,\,\, IJ^\pi=\frac l2\left(\frac l2 \pm \frac 12\right)^\pi
\ee
with $n_z=0,1,2, ..$ counting the number of quanta associated to the collective motion in the
holographic direction, and $n_\rho=0,1,2,..$ counting the number of quanta associated to the
radial breathing of the instanton core, a sort of Roper-like excitations. Following~\cite{SSXB}, 
we identify the parity of the heavy  baryon bound state  as $(-1)^{n_z}$.
Using (\ref{RX25}) and the results in~\cite{SSXB} as briefly summarized in the Appendix,  the mass 
spectrum for the bound heavy-light states is

\begin{eqnarray}
\label{RX27}
&&M_{NQ}=+M_0 +N_Qm_H\nonumber\\
&&+\left(\frac{(l+1)^2}6+\frac 2{15}N_c^2\left(1-\frac {15N_Q}{4 N_c}+\frac{5N_Q^2}{3N_c^2}\right)\right)^{\frac 12}M_{KK}\nonumber\\
&&+ \frac{2(n_\rho+n_z)+2}{\sqrt{6}}M_{KK}
\end{eqnarray}
with $M_{KK}$ the Kaluza-Klein mass and $M_0/M_{KK}=8\pi^2\kappa$ the bulk instanton mass. 
The Kaluza-Klein scale is usually set by the light meson spectrum and is fit to reproduce the rho 
mass with $M_{KK}\sim m_\rho/\sqrt{0.61}\sim 1$ GeV~\cite{SSX}. Whenever possible, we will try
to eliminate the uncertainties on the value of $M_{KK}$ through model independent relations for
fixed $N_Q$. 

We note that the net effect of the heavy-meson is among other thinghs,
an increase in the iso-rotational inertia by  expanding (\ref{RX27}) in $1/N_c$. 
The negative  $N_Q/N_c$ contribution in (\ref{RX27})  reflects on the fact that a heavy
meson with a heavy quark mass is attracted to the instanton to order $\lambda^0$. 
As we noted earlier, a heavy meson with a heavy anti-quark will be repelled to order
$\lambda^0$ hence a similar but positive contribution. The
positive $N_Q^2/N_c^2$ contribution is the repulsive Coulomb-like self-interaction. Note that
it is of the same order as the rotational contribution which justifies keeping it in our analysis.

(\ref{RX27}) is to be contrasted with the mass spectrum for baryons with no heavy quarks or $N_Q=0$, where 
the nucleon state is idendified as $N_Q=0,l=1,n_z=n_\rho=0$ and the Delta state as $N_Q=0,l=3,n_z=n_\rho=0$~\cite{SSXB}. The radial
excitation with $n_\rho=1$ can be identified with the radial Roper excitation of the nucleon and Delta, while the holographic excitation
with $n_z=1$ can be interpreted as the odd parity excitation of the nucleon and Delta.

\subsection{Single-heavy baryons}

Since the bound zero-mode transmuted to spin $\frac 12$, 
the lowest heavy baryons with one heavy quark are characterized 
by $N_Q=1, l=even,N_c=3$ and $n_z,n_\rho=0,1$, with the mass
spectrum

\begin{eqnarray}
\label{RX28}
M_{X_Q}=&&+M_0+m_H\\&&+\left(\frac{(l+1)^2}{6}
-\frac{7}{90}\right)^{\frac 12}M_{KK}\nonumber\\
&&+\frac {2(n_\rho+n_z)+2}{\sqrt{6}} M_{KK}
\end{eqnarray}

\subsubsection{Heavy baryons}

Consider the states with $n_z=n_\rho=0$. 
We identify the state with $l=0$ with the heavy-light iso-singlet $\Lambda_{Q}$ with the assignments
$IJ^\pi=0{\frac{1}{2}}^{+}$. We identify the state with $l=2$ with the heavy-light iso-triplet
$\Sigma_{Q}$ with the assignment $1{\frac{1}{2}}^+$, and $\Sigma_{Q}^{\star}$ with 
the assignment $1{\frac{3}{2}}^+$. By subtracting the nucleon mass from (\ref{RX28}) we have

\begin{eqnarray}
\label{RX29}
M_{\Lambda_{Q}}-M_{N}-m_{H}=-1.06\,M_{KK}\nonumber\\
M_{\Sigma_{Q}}-M_{N}-m_{H}=-0.17\,M_{KK}\nonumber\\
M_{\Sigma^*_{Q}}-M_{N}-m_{H}= -0.17\,M_{KK}
\end{eqnarray}
Hence the holographic and model independent relations

\be
\label{RX29X0}
&&M_{\Lambda_{Q^\prime}}= M_{\Lambda_{Q}} +(m_{H^\prime}-m_H)\nonumber\\
&&M_{\Sigma_{Q^\prime}}=0.84\,m_N+m_{H^{\prime}} +0.16\,(M_{\Lambda_{Q}}-m_H)
\ee
%\frac{M_{\Lambda_{Q}}-M_{N}-m_{Q}}{M_{\Sigma_{Q^\prime}}-M_{N}-m_{Q^\prime}}=\frac {0.326}{0.683}\sim \frac 12
%\ee
with $Q,Q^\prime=c,b$. Using the heavy meson masses $m_D\approx 1870$ Mev,  
$m_B=5279$ MeV and $m_{\Lambda_c}=2286$ Mev we find that
$M_{\Lambda_{b}}= 5655$ MeV in good agreement with the measured value of $5620$ MeV. 
Also we find $M_{\Sigma c}=2725$ Mev and $M_{\Sigma b}=6134$ Mev, which are to be compared
to the empirical values of $M_{\Sigma c}=2453$ Mev and $M_{\Sigma b}=5810$ Mev respectively.

\subsubsection{Excited heavy baryons}

Now, consider the low-lying breathing modes $R$ with $n_\rho=1$ for  the even assignments 
$0\frac 12^{+}, 1\frac 12^+, 1\frac 32^+$, and the odd parity excited states $O$ with $n_z=1$  
for the odd assigments $0\frac 12^{-}, 1\frac 12^-, 1\frac 32^-$. (\ref{RX28}) shows that the R-excitations
are degenerate with the O-excitations.  We
obtain ($E=O,R$)

\begin{eqnarray}
\label{RX29X1}
&&M_{\Lambda_{EQ^\prime}}=+0.23\,M_{\Lambda_{Q}}+0.77\,m_N-0.23\,m_H+m_{H^{\prime}}\nonumber\\
&&M_{\Sigma_{EQ^\prime}}=-0.59\,M_{\Lambda_{Q}}+1.59\,m_N+0.59\,m_H+m_{H^{\prime}}\nonumber\\
\end{eqnarray}
We found $M_{\Lambda_{Oc}}=2686$ MeV which is to be compared
to the mass  $2595$ MeV for the reported charm $0\frac 12^-$ state, and  $M_{\Lambda_{Ob}}=6095$ MeV which
is close to the mass $5912$ MeV for the reported bottom $0\frac 12^-$ state.  (\ref{RX29X1}) predicts 
a mass of $M_{\Sigma_{Oc}}=3126$ MeV for a possible charm $1\frac 12^-$ state, and a mass of
$M_{\Sigma_{Ob}}=6535$ MeV for a possible bottom $1\frac 12^-$ state.

\subsection{Double-heavy baryons}

For heavy baryons containing also anti-heavy quarks we note that a rerun of the preceding arguments 
using instead the reduction $\Phi_M=\phi_Me^{+im_Hx_0}$, amounts to binding an anti-heavy-light meson to 
the bulk instanton in the form of  a zero-mode also in the fundamental representation of spin. Most
of the results are unchanged except for pertinent minus signs. For instance, when binding one heavy-light
and one anti-heavy-light meson, (\ref{RX20}) now reads

\begin{eqnarray}
\label{RX30}
{\cal L}=&&+{\cal L}_0[a_I, X_\alpha]\nonumber\\
&&+\chi_Q^{\dagger}i\partial_t\chi_Q +\frac{3}{32\pi^2a\rho^2}\chi_Q^{\dagger}\chi_Q\nonumber\\
&&-\chi_{\bar Q}^{\dagger}i\partial_t\chi_{\bar Q} -\frac{3}{32\pi^2a\rho^2}\chi_{\bar Q}^{\dagger}\chi_{\bar Q}\nonumber\\
&&+\frac{(\chi_Q^{\dagger}\chi_Q-\chi_{\bar Q}^{\dagger}\chi_{\bar Q})^2}{24\pi^2a \rho^2N_c}
\end{eqnarray}
As we indicated earlier the mass contributions are opposite for a heavy-light and anti-heavy-light meson.
The general mass spectrum for baryons
with $N_Q$ heavy-quarks and $N_{\bar Q}$ anti-heavy quarks is

\begin{eqnarray}
\label{RX27X}
&&M_{\bar Q Q}=+M_0 +(N_{Q}+N_{\bar Q})m_H\nonumber\\
&&+\left(\frac{(l+1)^2}6\right.\nonumber\\
&&\left.+\frac 2{15}N_c^2\left(1-\frac{15(N_Q-N_{\bar Q})}{4N_c}+
\frac{5(N_Q-N_{\bar Q})^2}{3N_c^2}\right)\right)^{\frac 12}M_{KK}\nonumber\\
&&+ \frac{2(n_\rho+n_z)+2}{\sqrt{6}}M_{KK}
\end{eqnarray} 
%The double-heavy 
%baryons with $QQ$ content are heavier because of the larger U(1)  Coulomb-like repulsion in (\ref{RX27X}).

\subsubsection{Pentaquarks}

For $N_Q=N_{\bar Q}=1$
we identify the lowest state with $l=1, n_z=n_\rho=0$ with pentaquark baryonic states with the $IJ^\pi$ assignments 
$\frac 12 {\frac 12}^-$ and $\frac 12 {\frac 32}^-$, and masses given by

\begin{eqnarray}
\label{RX310}
M_{\bar QQ}-M_{N}-2\,m_{H}=0
%M_{\bar Q Q}=m_{N}+2m_{H}-0.140(M_{\Lambda_Q^{\prime}}-m_{N}-m_{H^{\prime}})
\end{eqnarray}
Amusingly the spectrum is BPS as both the attraction and repulsion balances, and the two
Coulomb-like self repulsions balance  against the  Coulomb-like  pair attraction.
Thus we predict a mass of $M_{\bar cc}=4678$ MeV for the $\frac{1}{2}\frac{3}{2}^-$ which is close to the 
reported  $P_c^+(4380)$ and $P_c^+(4450)$. We also predict a mass of
 $M_{\bar b c}=8087$ MeV and $M_{\bar bb}=11496$ MeV for the yet to be oberved pentaquarks. 
 Perhaps a better estimate for the latters is to trade $M_N$ in (\ref{RX31}) for the observed light 
 charmed pentaquark mass  $M_{\bar cc}=4678$ MeV using instead
 
 \be
 \label{RX311}
 M_{\bar QQ}=M_{\bar Q^\prime Q^\prime} +2\,(m_H-m_{H^\prime})
 \ee
 Using (\ref{RX311}) we predict $M_{\bar b c}=7789$ MeV and $M_{\bar bb}=11198$ MeV,
 which are slightly lighter than the previous estimates.
The present holographic construction based on the bulk instanton
as a hedgehog in flavor-spin space does not support the $\frac 12\frac 52^+$ assignment suggested for the 
observed $P_c^+(4450)$ through the bound zero-mode for the case $N_f=2$.
% It maybe allowed for a larger
%number of $N_f$ as we will addressed elsewhere.
%only. This higher spin and even parity assignment 
%may only be accomodated through a non-collective quantum excitation of the pion or vector field. This
%quantum analysis goes beyond the scope of this work. 

\subsubsection{Excited pentaquarks}

For $N_Q=N_{\bar Q}=1$ we now identify the lowest state with $l=1, n_z=1, n_\rho=0$ with the odd 
parity pentaquarks $O$ with assignments $\frac 12 {\frac 12}^+$ and $\frac 12 {\frac 32}^+$, and the $l=1, n_z=0, n_\rho=1$
with the breathing or Roper $R$ pentaquarks with the same assignments as the ground state.
The mass relations for these states are ($E=O,R$)

\begin{eqnarray}
\label{RX32}
M_{E\bar QQ}-M_{N}-2\,m_{H}=0.82\,M_{KK}
%M_{E \bar Q Q}=m_{N}+2m_{H}-0.91(M_{\Lambda_Q^{\prime}}-m_{N}-m_{H^{\prime}})
\end{eqnarray}
which can be traded for model independent relations

\begin{eqnarray}
\label{RX322}
%M_{E\bar QQ}-M_{N}-2\,m_{H}=0.82\,M_{KK}
M_{E \bar Q Q}=1.51\, m_{N}+2m_{H}+0.51\,(m_{H^\prime}-M_{\lambda_{Q^\prime}})
\end{eqnarray}
by eliminating $M_{KK}$ using the first relation in  (\ref{RX29}).  
Using (\ref{RX322}) we predict $M_{E\bar cc}=4944$ MeV, $M_{E\bar bc}=8353$ MeV, $M_{E\bar bb}=11762$ MeV
as the new low lying excitations of heavy pentaquarks with the preceding assignments. 

\subsubsection{Delta-like pentaquarks}

For $N_Q=N_{\bar Q}=1$, the present construction allows also for Delta-type pentaquarks which 
we identify with  $l=3, n_z=n_\rho=0$. Altogether, we have one $\frac 32 \frac 12^-$, two $\frac 32\frac 32^-$, 
and one $\frac 32 \frac 52^-$ states, all degenerate to leading order,  with heavy flavor dependent 
masses

\begin{eqnarray}
\label{RX31}
M_{\Delta\bar QQ}-M_{N}-2\,m_{Q}=0.71\,M_{KK}
%M_{\Delta\bar QQ}=2m_{H}+m_N-0.65(M_{\Lambda_Q^{\prime}}-m_{N}-m_{H^{\prime}})
\end{eqnarray}
Again we can trade $M_{KK}$ using the first relation in  (\ref{RX29}) to obtain the model
independent relation

\begin{eqnarray}
\label{RX31X}
%M_{\Delta\bar QQ}-M_{N}-2\,m_{Q}=0.71\,M_{KK}
M_{\Delta\bar QQ}=1.57\,m_N+2\,m_{H}+0.57\,(m_{H^\prime}-M_{\Lambda_{Q^\prime}})
\end{eqnarray}
In particular, we predict $M_{\Delta\bar c c}=4976$ MeV, $M_{\Delta\bar c b}=8385$ MeV, 
and $M_{\Delta\bar b b}=11794$ MeV, which are yet to be observed.

%%%%%%%%%%%%%%%%%%%%%%%%%%%%%%%%%%%%%%%%%%%%%%%%%%%%%%%%%%%%%%%%%%%

%%%%%%%%%%%%%%%%%%%%%%%%%%%%%%%%%%%%%%%%%%%%%%%%%%%%%%%%%%%%%%%%%%%

\section{Conclusions}

We have presented a top-down holographic approach to the single- and double-heavy baryons 
in the variant of $D4$-$D8$ we proposed recently~\cite{LIUHEAVY} (first reference).  To order
$\lambda m_Q^0$, the heavy baryons emerge from the zero mode of a reduced (massless)
vector meson that transmutes both its spin and negative parity,
to a spin $\frac 12$ with positive parity in the bulk flavor instanton. Heavy mesons and
anti-mesons bind on equal footing to the core instanton in holography in leading order
in $\lambda$ even in the presence of
the Chern-Simons contribution. This is not the case in  non-holographic models where the 
anti-heavy meson binding is usually depressed by the sign flip in the Wess-Zumino-Witten
contribution~\cite{SKYRMEHEAVY}. Unlike in the Skyrme model, the bulk flavor instanton 
offers a model independent description of the light baryon core. The binding of the
heavy meson over its Compton wavelength is essentially geometrical in the double limit
of large $\lambda$ followed by large $m_Q$.

We have shown that the  bound state moduli yields a rich spectrum after quantization,
 that involves coupled rotational, translational and vibrational modes. The model-independent
 mass relations for the low-lying  single-heavy baryon spectrum yield masses that are in 
  overall agreement with the reported masses for the corresponding charm and bottom 
  baryons.  The spectrum also contains some newly excited states yet to be observed. 
When extended to double-heavy baryon spectra, the holographic contruction yields 
a pair of degenerate heavy iso-doublets with  $IJ^\pi=\frac 12 {\frac 12}^-, \frac 12 {\frac 32}^-$
assignments.  The model gives naturally a charmed pentaquark . It also predicts a number of new pentaquarks 
with both hidden charm and bottom, and five new Delta-like pentaquarks with hidden charm. 
The hedgehog flavor instanton when collectively quantized, excludes the $IJ^\pi=\frac 12 {\frac 52}^+$ assignment
for $N_f=2$.

 The shortcomings of the heavy-light holographic approach stem from the triple limits of large $N_c$ and strong 
 $^\prime$t Hooft coupling $\lambda=g^2N_c$, and now  large $m_H$  as well. 
 The corrections are clear in principle but laborious in practice. Our simple construct 
can be improved through a more realistic extension such as improved holographic 
QCD~\cite{KIRITSIS}.  Also a simpler, bottom-up formulation following the present general reasoning
is also worth formulating for the transparency of the arguments.

Finally, it would be interesting to extend the current analysis for the heavy baryons
to the more realistic case of
$N_f=3$ with a realistic mass for the light strange quark as well. Also, 
the strong decay widths of the heavy baryons and their exotics should be estimated.
They follow from $1/N_c$ type corrections using the self-generated Yukawa-type
potentials in bulk, much like those studied in the context of the Skyrme model~\cite{ADAMI}.
We expect large widths to develop through S-wave decays, and smaller  widths
to follow from P-wave decays because of a smaller phase space. Also the hyperfine splitting
in the heavy spectra is expected to arise through subleading couplings between the emerging spin
degrees of freedom and the collective rotations and vibrations. The pertinent electromagnetic
and weak form factors of the holographically bound heavy baryons can also be obtained following
standard arguments~\cite{SSXB,SSXBB}. Some of these issues will be addressed next.

\section{Acknowledgements}

We thank Rene Meyer for a discussion. 
This work was supported by the U.S. Department of Energy under Contract No.
DE-FG-88ER40388.

\section{Appendix}

In this Appendix we summarize some of the essential steps for the quantization of the instanton
moduli developed in~\cite{SSXB}, and fill up for  some of the notations used in the main
text. In the absence of the
heavy mesons, we also take the   large $\lambda$ limit using  the same rescaling 
to re-write the contributions of the light gauge fields as 

\be
S=aN_c \lambda S_{YM}(A_{M},\hat A_{M})+aN_cS_1(A_0,\hat A_0,A_{M},\hat A_{M})
\ee 
Here $A$ refers to  the SU(2) part of the light gauge field, and 
$\hat A$ to its U(1) part. The equation of motion for $A_M,\hat A_M$ are at leading order of $\lambda$ 

\be
D_{N}F_{NM}=0\qquad and \qquad
\partial_N \hat F_{NM}=0
\ee
They are solved using the flat instanton 
$A_M$ and 0 for $\hat A_M$. The  equation of motion for the  time components are subleading

\begin{eqnarray}
&&D_{M}F_{0M}+\frac{1}{64\pi^2a}\epsilon_{MNPQ}\hat F_{MN}F_{PQ}=0\nonumber\\
&&\partial_{M}\hat F_{0M}+\frac{1}{64\pi^2a}\epsilon_{MNPQ}tr F_{MN}F_{PQ}=0
\end{eqnarray}
They are solved using $A_0=0$ and a non-zero $\hat A_0$ as defined in the main text.

To obtain the spectrum we promote the moduli of the solution to be time dependent, i.e.

\be
({ a}_I, X_{\alpha})\rightarrow ({a}_I(t), X_{\alpha}(t))
\ee
Here ${a}_I$ refers to the moduli of the global SU(2) gauge transformation. 
In order to satisfy the constraint equation (52) (Gauss's law) we need to impose a further gauge transformation on
the field configuation 

\begin{eqnarray}
A^V_{M}=V^{\dagger}(A_{M}+\partial_{M})V\qquad and \qquad
A^V_{0}=V^{\dagger}\partial_{t}	V
\end{eqnarray}
Inserting  the transformed field configuration in the constraint equation, we find that $V$ is solved by

\begin{eqnarray}
-iV^{\dagger}\partial_{t}	V=\Phi=-\partial_t X_{N}A_{N}+\chi_a \Phi_a
%\nonumber\\
%&&\chi_a=-iTr(\tau^{a}{\bf a}^{-1}\partial_t {\bf a})
\end{eqnarray}
with $\chi_a[{ a}_I]$ as defined in the main text.
Putting the resulting slowly moving field configuration back in the action, allows for
the light collective Hamiltonian~\cite{SSXB}

\begin{eqnarray}
&&H_0=M_0+H_{Z}+H_{\rho}\nonumber\\
&&H_{Z}=-\frac{\partial_Z^2}{2m_z}+\frac{m_z\omega_z^2}{2} Z^2\nonumber\\
&&H_{\rho}=-\frac{\nabla_y^2}{2m_y}+\frac{m_y\omega_\rho^2}{2} \rho^2+\frac{Q}{\rho^2}\nonumber\\
&&y=\rho(a_1,a_2,a_3,a_4),{a}_I=a_4+i\vec{a}\cdot\vec \tau\nonumber\\
&&m_z=\frac{m_y}{2}=8\pi^2 aN_c, \omega_z^2=\frac{2}{3}, \omega_\rho^2=\frac{1}{6}
\end{eqnarray}
The eigenstates of  $H_\rho$ are given by $T^{l}(a)R_{l,n_\rho}(\rho)$, where $T^{l}$ are the spherical harmonics on $S^3$.
Under $SO(4)=SU(2)\times SU(2)/Z_2$ they  are in the $(\frac{l}{2},\frac{l}{2})$ representations, where the two SU(2) factors are defined by the isometry ${a}_I\rightarrow V_L {a}_IV_{R}$.  The left factor is the isospin rotation, and the right factor is the  space rotation. This quantization describes ${I}=J=\frac{l}{2}$ states. The nucleon is realized as the lowest state with $l=1$ and $n_{\rho}=n_z=0$.

%%%%%%%%%%%%%%%%%%%%%%%

 \vfil


\begin{thebibliography}{99} \frenchspacing


\bibitem{ISGUR}
%\cite{Shuryak:1981fza}
%\bibitem{Shuryak:1981fza} 
  E.~V.~Shuryak,
  %``Hadrons Containing a Heavy Quark and QCD Sum Rules,''
  Nucl.\ Phys.\ B {\bf 198}, 83 (1982);
 % doi:10.1016/0550-3213(82)90546-6
  %%CITATION = doi:10.1016/0550-3213(82)90546-6;%%
  %370 citations counted in INSPIRE as of 10 Apr 2017
%\cite{Isgur:1991wq}
%\bibitem{Isgur:1991wq}
  N.~Isgur and M.~B.~Wise,
  %``Spectroscopy with heavy quark symmetry,''
  Phys.\ Rev.\ Lett.\  {\bf 66} (1991) 1130;
 % doi:10.1103/PhysRevLett.66.1130
  %%CITATION = doi:10.1103/PhysRevLett.66.1130;%%
  %456 citations counted in INSPIRE as of 19 Aug 2016
%\cite{Manohar:2000dt}
%\bibitem{Manohar:2000dt} 
  A.~V.~Manohar and M.~B.~Wise,
  ``Heavy quark physics,''
  Camb.\ Monogr.\ Part.\ Phys.\ Nucl.\ Phys.\ Cosmol.\  {\bf 10}, 1 (2000).
  %%CITATION = CMPCE,10,1;%%
  %302 citations counted in INSPIRE as of 19 Aug 2016


\bibitem{MACIEK}
%\cite{Nowak:1992um}
%\bibitem{Nowak:1992um} 
  M.~A.~Nowak, M.~Rho and I.~Zahed,
  %``Chiral effective action with heavy quark symmetry,''
  Phys.\ Rev.\ D {\bf 48}, 4370 (1993)
  %doi:10.1103/PhysRevD.48.4370
  [hep-ph/9209272];
  %%CITATION = doi:10.1103/PhysRevD.48.4370;%%
  %160 citations counted in INSPIRE as of 28 Jun 2016
%\cite{Nowak:2003ra}
%\bibitem{Nowak:2003ra} 
  M.~A.~Nowak, M.~Rho and I.~Zahed,
  %``Chiral doubling of heavy light hadrons: BABAR 2317-MeV/c**2 and CLEO 2463-MeV/c**2 discoveries,''
  Acta Phys.\ Polon.\ B {\bf 35}, 2377 (2004)
  [hep-ph/0307102].
  %%CITATION = HEP-PH/0307102;%%
  %131 citations counted in INSPIRE as of 28 Jun 2016


\bibitem{BARDEEN}
%\cite{Bardeen:1993ae}
%\bibitem{Bardeen:1993ae}
  W.~A.~Bardeen and C.~T.~Hill,
  %``Chiral dynamics and heavy quark symmetry in a solvable toy field theoretic model,''
  Phys.\ Rev.\ D {\bf 49} (1994) 409
  %doi:10.1103/PhysRevD.49.409
  [hep-ph/9304265];
  %%CITATION = doi:10.1103/PhysRevD.49.409;%%
  %207 citations counted in INSPIRE as of 28 Jun 2016
%\cite{Bardeen:2003kt}
%\bibitem{Bardeen:2003kt} 
  W.~A.~Bardeen, E.~J.~Eichten and C.~T.~Hill,
  %``Chiral multiplets of heavy - light mesons,''
  Phys.\ Rev.\ D {\bf 68}, 054024 (2003)
 % doi:10.1103/PhysRevD.68.054024
  [hep-ph/0305049].
  %%CITATION = doi:10.1103/PhysRevD.68.054024;%%
  %399 citations counted in INSPIRE as of 28 Jun 2016




   
\bibitem{BABAR}
%\cite{Aubert:2003fg}
%\bibitem{Aubert:2003fg} 
  B.~Aubert {\it et al.} [BaBar Collaboration],
  %``Observation of a narrow meson decaying to $D_s^+ \pi^0$ at a mass of 2.32-GeV/c$^2$,''
  Phys.\ Rev.\ Lett.\  {\bf 90}, 242001 (2003)
  %doi:10.1103/PhysRevLett.90.242001
  [hep-ex/0304021].
  %%CITATION = doi:10.1103/PhysRevLett.90.242001;%%
  %752 citations counted in INSPIRE as of 28 Jun 2016

\bibitem{CLEOII}
%\cite{Besson:2003cp}
%\bibitem{Besson:2003cp} 
  D.~Besson {\it et al.} [CLEO Collaboration],
  %``Observation of a narrow resonance of mass 2.46-GeV/c**2 decaying to D*+(s) pi0 and confirmation of the D*(sJ)(2317) state,''
  Phys.\ Rev.\ D {\bf 68}, 032002 (2003)
  Erratum: [Phys.\ Rev.\ D {\bf 75}, 119908 (2007)]
 % doi:10.1103/PhysRevD.68.032002, 10.1103/PhysRevD.75.119908
  [hep-ex/0305100].
  %%CITATION = doi:10.1103/PhysRevD.68.032002, 10.1103/PhysRevD.75.119908;%%
  %525 citations counted in INSPIRE as of 28 Jun 2016







\bibitem{BELLE}
%\cite{Collaboration:2011gja}
%\bibitem{Collaboration:2011gja} 
  I.~Adachi [Belle Collaboration],
  %``Observation of two charged bottomonium-like resonances,''
  arXiv:1105.4583 [hep-ex];
  %%CITATION = ARXIV:1105.4583;%%
  %145 citations counted in INSPIRE as of 28 Jun 2016
%\cite{Belle:2011aa}
%\bibitem{Belle:2011aa} 
  A.~Bondar {\it et al.} [Belle Collaboration],
  %``Observation of two charged bottomonium-like resonances in Y(5S) decays,''
  Phys.\ Rev.\ Lett.\  {\bf 108}, 122001 (2012)
  %doi:10.1103/PhysRevLett.108.122001
  [arXiv:1110.2251 [hep-ex]].
  %%CITATION = doi:10.1103/PhysRevLett.108.122001;%%
  %291 citations counted in INSPIRE as of 28 Jun 2016


\bibitem{BESIII}
%\cite{Ablikim:2013mio}
%\bibitem{Ablikim:2013mio} 
  M.~Ablikim {\it et al.} [BESIII Collaboration],
  %``Observation of a Charged Charmoniumlike Structure in $e^+e^-$ ? J/?  at $\sqrt{s}$ =4.26??GeV,''
  Phys.\ Rev.\ Lett.\  {\bf 110}, 252001 (2013)
  %doi:10.1103/PhysRevLett.110.252001
  [arXiv:1303.5949 [hep-ex]].
  %%CITATION = doi:10.1103/PhysRevLett.110.252001;%%
  %384 citations counted in INSPIRE as of 28 Jun 2016



\bibitem{DO}
%\cite{D0:2016mwd}
%\bibitem{D0:2016mwd}
  V.~M.~Abazov {\it et al.} [D0 Collaboration],
  %``Evidence for a $B_s^0 \pi^\pm$ state,''
  %Submitted to: Phys.Rev.Lett.
  [arXiv:1602.07588 [hep-ex]].
  %%CITATION = ARXIV:1602.07588;%%
  %42 citations counted in INSPIRE as of 28 Jun 2016

\bibitem{LHCb}
%\cite{Aaij:2016iza}
%\bibitem{Aaij:2016iza}
  R.~Aaij {\it et al.} [LHCb Collaboration],
  %``Observation of $J/\psi\phi$ structures consistent with exotic states from amplitude analysis of $B^+\to J/\psi \phi K^+$ decays,''
  arXiv:1606.07895 [hep-ex];
  %%CITATION = ARXIV:1606.07895;%%
%\cite{Aaij:2016nsc}
%\bibitem{Aaij:2016nsc}
  R.~Aaij {\it et al.} [LHCb Collaboration],
  %``Amplitude analysis of $B^+\to J/\psi \phi K^+$ decays,''
  arXiv:1606.07898 [hep-ex].
  %%CITATION = ARXIV:1606.07898;%%
  
  
  \bibitem{LHCbx}
  %\cite{Aaij:2015tga}
%\bibitem{Aaij:2015tga}
  R.~Aaij {\it et al.} [LHCb Collaboration],
  %``Observation of $J/\psi p$ Resonances Consistent with Pentaquark States in $\Lambda_b^0 \to J/\psi K^- p$ Decays,''
  Phys.\ Rev.\ Lett.\  {\bf 115} (2015) 072001
 % doi:10.1103/PhysRevLett.115.072001
  [arXiv:1507.03414 [hep-ex]];
  %%CITATION = doi:10.1103/PhysRevLett.115.072001;%%
  %367 citations counted in INSPIRE as of 05 Apr 2017
%\cite{Aaij:2016phn}
%\bibitem{Aaij:2016phn}
  R.~Aaij {\it et al.} [LHCb Collaboration],
  %``Model-independent evidence for $J/\psi p$ contributions to $\Lambda_b^0\to J/\psi p K^-$ decays,''
  Phys.\ Rev.\ Lett.\  {\bf 117} (2016) no.8,  082002
 % doi:10.1103/PhysRevLett.117.082002
  [arXiv:1604.05708 [hep-ex]];
  %%CITATION = doi:10.1103/PhysRevLett.117.082002;%%
  %31 citations counted in INSPIRE as of 05 Apr 2017
%\cite{Aaij:2016ymb}
%\bibitem{Aaij:2016ymb}
  R.~Aaij {\it et al.} [LHCb Collaboration],
  %``Evidence for exotic hadron contributions to $\Lambda_b^0 \to J/\psi p \pi^-$ decays,''
  Phys.\ Rev.\ Lett.\  {\bf 117} (2016) no.8,  082003
   Addendum: [Phys.\ Rev.\ Lett.\  {\bf 117} (2016) no.10,  109902]
  %doi:10.1103/PhysRevLett.118.119901, 10.1103/PhysRevLett.117.082003, 10.1103/PhysRevLett.117.109902
  [arXiv:1606.06999 [hep-ex]].
  %%CITATION = doi:10.1103/PhysRevLett.118.119901, 10.1103/PhysRevLett.117.082003, 10.1103/PhysRevLett.117.109902;%%
  %32 citations counted in INSPIRE as of 05 Apr 2017


\bibitem{LHCbxx}
%\cite{Aaij:2017nav}
%\bibitem{Aaij:2017nav} 
  R.~Aaij {\it et al.} [LHCb Collaboration],
  %``Observation of five new narrow $\Omega_c^0$ states decaying to $\Xi_c^+ K^-$,''
  arXiv:1703.04639 [hep-ex].
  %%CITATION = ARXIV:1703.04639;%%
  %8 citations counted in INSPIRE as of 05 Apr 2017




\bibitem{MOLECULES}
%\cite{Voloshin:1976ap}
%\bibitem{Voloshin:1976ap} 
  M.~B.~Voloshin and L.~B.~Okun,
  %``Hadron Molecules and Charmonium Atom,''
  JETP Lett.\  {\bf 23}, 333 (1976)
  [Pisma Zh.\ Eksp.\ Teor.\ Fiz.\  {\bf 23}, 369 (1976)];
  %%CITATION = JTPLA,23,333;%%
  %292 citations counted in INSPIRE as of 28 Jun 2016

\bibitem{THORSSON}
  %\cite{Tornqvist:1991ks}
%\bibitem{Tornqvist:1991ks} 
  N.~A.~Tornqvist,
  %``Possible large deuteron - like meson meson states bound by pions,''
  Phys.\ Rev.\ Lett.\  {\bf 67}, 556 (1991);
  %doi:10.1103/PhysRevLett.67.556
  %%CITATION = doi:10.1103/PhysRevLett.67.556;%%
  %232 citations counted in INSPIRE as of 28 Jun 2016
%\cite{Tornqvist:1993ng}
%\bibitem{Tornqvist:1993ng} 
  N.~A.~Tornqvist,
  %``From the deuteron to deusons, an analysis of deuteron - like meson meson bound states,''
  Z.\ Phys.\ C {\bf 61}, 525 (1994)
 % doi:10.1007/BF01413192
  [hep-ph/9310247];
  %%CITATION = doi:10.1007/BF01413192;%%
  %281 citations counted in INSPIRE as of 15 Jul 2016
%\cite{Tornqvist:2004qy}
%\bibitem{Tornqvist:2004qy} 
  N.~A.~Tornqvist,
  %``Isospin breaking of the narrow charmonium state of Belle at 3872-MeV as a deuson,''
  Phys.\ Lett.\ B {\bf 590}, 209 (2004)
%  doi:10.1016/j.physletb.2004.03.077
  [hep-ph/0402237].
  %%CITATION = doi:10.1016/j.physletb.2004.03.077;%%
  %424 citations counted in INSPIRE as of 15 Jul 2016



  

  
  
\bibitem{KARLINER}
%\cite{Karliner:2008rc}
%\bibitem{Karliner:2008rc} 
  M.~Karliner and H.~J.~Lipkin,
  %``Possibility of Exotic States in the Upsilon system,''
  arXiv:0802.0649 [hep-ph];
  %%CITATION = ARXIV:0802.0649;%%
  %33 citations counted in INSPIRE as of 04 Aug 2016
%\cite{Karliner:2015ina}
%\bibitem{Karliner:2015ina}
  M.~Karliner and J.~L.~Rosner,
  %``New Exotic Meson and Baryon Resonances from Doubly-Heavy Hadronic Molecules,''
  Phys.\ Rev.\ Lett.\  {\bf 115} (2015) no.12,  122001
  %doi:10.1103/PhysRevLett.115.122001
  [arXiv:1506.06386 [hep-ph]];
  %%CITATION = doi:10.1103/PhysRevLett.115.122001;%%
  %47 citations counted in INSPIRE as of 28 Jun 2016
%\cite{Karliner:2016via}
%\bibitem{Karliner:2016via} 
  M.~Karliner,
  %``Doubly Heavy Exotic Mesons and Baryons and How to Look for Them,''
  Acta Phys.\ Polon.\ B {\bf 47}, 117 (2016).
  %doi:10.5506/APhysPolB.47.117
  %%CITATION = doi:10.5506/APhysPolB.47.117;%%
  %1 citations counted in INSPIRE as of 28 Jun 2016



\bibitem{OTHERS}
%\cite{Thomas:2008ja}
%\bibitem{Thomas:2008ja} 
  C.~E.~Thomas and F.~E.~Close,
  %``Is X(3872) a molecule?,''
  Phys.\ Rev.\ D {\bf 78}, 034007 (2008)
  %doi:10.1103/PhysRevD.78.034007
  [arXiv:0805.3653 [hep-ph]];
 %\cite{Close:2010wq}
%\bibitem{Close:2010wq} 
  F.~Close, C.~Downum and C.~E.~Thomas,
  %``Novel Charmonium and Bottomonium Spectroscopies due to Deeply Bound Hadronic Molecules from Single Pion Exchange,''
  Phys.\ Rev.\ D {\bf 81}, 074033 (2010)
  %doi:10.1103/PhysRevD.81.074033
  [arXiv:1001.2553 [hep-ph]].
  %%CITATION = doi:10.1103/PhysRevD.81.074033;%%
  %35 citations counted in INSPIRE as of 12 Jul 2016
 %%CITATION = doi:10.1103/PhysRevD.78.034007;%%
  %94 citations counted in INSPIRE as of 12 Jul 2016
%\cite{Ohkoda:2012hv}
%\bibitem{Ohkoda:2012hv} 


  \bibitem{OTHERSX}
  S.~Ohkoda, Y.~Yamaguchi, S.~Yasui, K.~Sudoh and A.~Hosaka,
  %``Exotic mesons with double charm and bottom flavor,''
  Phys.\ Rev.\ D {\bf 86}, 034019 (2012)
 % doi:10.1103/PhysRevD.86.034019
  [arXiv:1202.0760 [hep-ph]];
  %%CITATION = doi:10.1103/PhysRevD.86.034019;%%
  %17 citations counted in INSPIRE as of 12 Jul 2016
%\cite{Ohkoda:2012pt}
%\bibitem{Ohkoda:2012pt} 
  S.~Ohkoda, Y.~Yamaguchi, S.~Yasui, K.~Sudoh and A.~Hosaka,
  %``Exotic mesons with hidden charm and bottom near thresholds,''
  arXiv:1209.0144 [hep-ph].


\bibitem{OTHERSZ}
%\cite{AlFiky:2005jd}
%\bibitem{AlFiky:2005jd} 
  M.~T.~AlFiky, F.~Gabbiani and A.~A.~Petrov,
  %``X(3872): Hadronic molecules in effective field theory,''
  Phys.\ Lett.\ B {\bf 640}, 238 (2006)
 % doi:10.1016/j.physletb.2006.07.069
  [hep-ph/0506141];
  %%CITATION = doi:10.1016/j.physletb.2006.07.069;%%
  %102 citations counted in INSPIRE as of 25 Aug 2016
%\cite{Lee:2009hy}
%\bibitem{Lee:2009hy} 
  I.~W.~Lee, A.~Faessler, T.~Gutsche and V.~E.~Lyubovitskij,
  %``X(3872) as a molecular DD* state in a potential model,''
  Phys.\ Rev.\ D {\bf 80}, 094005 (2009)
  %doi:10.1103/PhysRevD.80.094005
  [arXiv:0910.1009 [hep-ph]];
  %%CITATION = doi:10.1103/PhysRevD.80.094005;%%
  %54 citations counted in INSPIRE as of 19 Aug 2016
 %\cite{Suzuki:2005ha}
%\bibitem{Suzuki:2005ha} 
  M.~Suzuki,
  %``The X(3872) boson: Molecule or charmonium,''
  Phys.\ Rev.\ D {\bf 72}, 114013 (2005)
 % doi:10.1103/PhysRevD.72.114013
  [hep-ph/0508258];
  %%CITATION = doi:10.1103/PhysRevD.72.114013;%%
  %159 citations counted in INSPIRE as of 19 Aug 2016
 %\cite{Zhang:2011jja}
%\bibitem{Zhang:2011jja} 
  J.~R.~Zhang, M.~Zhong and M.~Q.~Huang,
  %``Could $Z_{b}(10610)$ be a $B^{*}\bar{B}$ molecular state?,''
  Phys.\ Lett.\ B {\bf 704}, 312 (2011)
 % doi:10.1016/j.physletb.2011.09.039
  [arXiv:1105.5472 [hep-ph]];
  %%CITATION = doi:10.1016/j.physletb.2011.09.039;%%
  %47 citations counted in INSPIRE as of 19 Aug 2016
 %\cite{Bugg:2011jr}
%\bibitem{Bugg:2011jr} 
  D.~V.~Bugg,
  %``An Explanation of Belle states $Z_b(10610)$ and $Z_b(10650)$,''
  Europhys.\ Lett.\  {\bf 96}, 11002 (2011)
%  doi:10.1209/0295-5075/96/11002
  [arXiv:1105.5492 [hep-ph]];
  %%CITATION = doi:10.1209/0295-5075/96/11002;%%
  %59 citations counted in INSPIRE as of 19 Aug 2016
 %\cite{Nieves:2011vw}
%\bibitem{Nieves:2011vw} 
  J.~Nieves and M.~P.~Valderrama,
  %``Deriving the existence of $B\bar{B}^*$ bound states from the X(3872) and Heavy Quark Symmetry,''
  Phys.\ Rev.\ D {\bf 84}, 056015 (2011)
  %doi:10.1103/PhysRevD.84.056015
  [arXiv:1106.0600 [hep-ph]];
  %%CITATION = doi:10.1103/PhysRevD.84.056015;%%
  %46 citations counted in INSPIRE as of 19 Aug 2016
 %\cite{Cleven:2011gp}
%\bibitem{Cleven:2011gp} 
  M.~Cleven, F.~K.~Guo, C.~Hanhart and U.~G.~Meissner,
  %``Bound state nature of the exotic $Z_b$ states,''
  Eur.\ Phys.\ J.\ A {\bf 47}, 120 (2011)
%  doi:10.1140/epja/i2011-11120-6
  [arXiv:1107.0254 [hep-ph]];
  %%CITATION = doi:10.1140/epja/i2011-11120-6;%%
  %80 citations counted in INSPIRE as of 19 Aug 2016
 %\cite{Mehen:2011yh}
%\bibitem{Mehen:2011yh} 
  T.~Mehen and J.~W.~Powell,
  %``Heavy Quark Symmetry Predictions for Weakly Bound B-Meson Molecules,''
  Phys.\ Rev.\ D {\bf 84}, 114013 (2011)
 % doi:10.1103/PhysRevD.84.114013
  [arXiv:1109.3479 [hep-ph]];
  %%CITATION = doi:10.1103/PhysRevD.84.114013;%%
  %50 citations counted in INSPIRE as of 19 Aug 2016
 %\bibitem{Guo:2013sya} 
  F.~K.~Guo, C.~Hidalgo-Duque, J.~Nieves and M.~P.~Valderrama,
  %``Consequences of Heavy Quark Symmetries for Hadronic Molecules,''
  Phys.\ Rev.\ D {\bf 88}, 054007 (2013)
  %doi:10.1103/PhysRevD.88.054007
  [arXiv:1303.6608 [hep-ph]];
  %%CITATION = doi:10.1103/PhysRevD.88.054007;%%
  %105 citations counted in INSPIRE as of 25 Oct 2016
 %\cite{Wang:2013cya}
%\bibitem{Wang:2013cya} 
  Q.~Wang, C.~Hanhart and Q.~Zhao,
  %``Decoding the riddle of $Y(4260)$ and $Z_c(3900)$,''
  Phys.\ Rev.\ Lett.\  {\bf 111}, no. 13, 132003 (2013)
 % doi:10.1103/PhysRevLett.111.132003
  [arXiv:1303.6355 [hep-ph]];
  %%CITATION = doi:10.1103/PhysRevLett.111.132003;%%
  %135 citations counted in INSPIRE as of 25 Oct 2016
 %\cite{Guo:2014iya}
%\bibitem{Guo:2014iya}
  F.~K.~Guo, C.~Hanhart, Q.~Wang and Q.~Zhao,
  %``Could the near-threshold $XYZ$ states be simply kinematic effects?,''
  Phys.\ Rev.\ D {\bf 91} (2015) no.5,  051504
  %doi:10.1103/PhysRevD.91.051504
  [arXiv:1411.5584 [hep-ph]];
  %%CITATION = doi:10.1103/PhysRevD.91.051504;%%
  %39 citations counted in INSPIRE as of 25 Oct 2016
 %\cite{Kang:2016ezb}
%\bibitem{Kang:2016ezb}
  X.~W.~Kang, Z.~H.~Guo and J.~A.~Oller,
  %``General considerations on the nature of $Z_b(10610)$ and $Z_b(10650)$ from their pole positions,''
  Phys.\ Rev.\ D {\bf 94} (2016) no.1,  014012
  %doi:10.1103/PhysRevD.94.014012
  [arXiv:1603.05546 [hep-ph]];
  %%CITATION = doi:10.1103/PhysRevD.94.014012;%%
  %1 citations counted in INSPIRE as of 25 Aug 2016
 %\cite{Guo:2013sya}
%\cite{Kang:2016jxw}
%\bibitem{Kang:2016jxw} 
  X.~W.~Kang and J.~A.~Oller,
  %``Different pole structures in line shapes of the $X(3872)$,''
  arXiv:1612.08420 [hep-ph].
  %%CITATION = ARXIV:1612.08420;%%
  %1 citations counted in INSPIRE as of 12 Apr 2017


 
 \bibitem{OTHERSXX}
 %\cite{Swanson:2006st}
%\bibitem{Swanson:2006st} 
  E.~S.~Swanson,
  %``The New heavy mesons: A Status report,''
  Phys.\ Rept.\  {\bf 429}, 243 (2006)
  %doi:10.1016/j.physrep.2006.04.003
  [hep-ph/0601110];
  %%CITATION = doi:10.1016/j.physrep.2006.04.003;%%
  %586 citations counted in INSPIRE as of 19 Aug 2016
 %\cite{Sun:2011uh}
%\bibitem{Sun:2011uh} 
  Z.~F.~Sun, J.~He, X.~Liu, Z.~G.~Luo and S.~L.~Zhu,
  %``$Z_b(10610)^\pm$ and $Z_b(10650)^\pm$ as the $B^*\bar{B}$ and $B^*\bar{B}^{*}$ molecular states,''
  Phys.\ Rev.\ D {\bf 84}, 054002 (2011)
  %doi:10.1103/PhysRevD.84.054002
  [arXiv:1106.2968 [hep-ph]];
  %%CITATION = doi:10.1103/PhysRevD.84.054002;%%
  %94 citations counted in INSPIRE as of 19 Aug 2016
 
 
 \bibitem{LIUMOLECULE}
%\cite{Liu:2016kqx}
%\bibitem{Liu:2016kqx} 
  Y.~Liu and I.~Zahed,
  %``Heavy Exotic Molecules with Charm and Bottom,''
  Phys.\ Lett.\ B {\bf 762}, 362 (2016)
 % doi:10.1016/j.physletb.2016.09.045
  [arXiv:1608.06535 [hep-ph]];
  %%CITATION = doi:10.1016/j.physletb.2016.09.045;%%
  %2 citations counted in INSPIRE as of 31 Oct 2016
 %\cite{Liu:2016anz}
%\bibitem{Liu:2016anz} 
  Y.~Liu and I.~Zahed,
  %``Heavy Exotic Molecules,''
  Int.\ J.\ Mod.\ Phys.\ E {\bf 26},  1740017 (2017)
 % doi:10.1142/S0218301317400171
  [arXiv:1610.06543 [hep-ph]];
  %%CITATION = doi:10.1142/S0218301317400171;%%
  %2 citations counted in INSPIRE as of 11 Apr 2017
 %\cite{Liu:2016urz}
%\bibitem{Liu:2016urz} 
  Y.~Liu and I.~Zahed,
  %``Heavy-Light Mesons in Chiral AdS/QCD,''
  arXiv:1611.04400 [hep-ph].
  %%CITATION = ARXIV:1611.04400;%%
  %3 citations counted in INSPIRE as of 11 Apr 2017
 
 
 %\cite{Albaladejo:2015lob}
\bibitem{Albaladejo:2015lob} 
  M.~Albaladejo, F.~K.~Guo, C.~Hidalgo-Duque and J.~Nieves,
  %``$Z_c(3900)$: What has been really seen?,''
  Phys.\ Lett.\ B {\bf 755}, 337 (2016)
  doi:10.1016/j.physletb.2016.02.025
  [arXiv:1512.03638 [hep-ph]].
  %%CITATION = doi:10.1016/j.physletb.2016.02.025;%%
  %8 citations counted in INSPIRE as of 25 Oct 2016
 
 
 
 
 
\bibitem{MANOHAR}
%\cite{Manohar:1992nd}
%\bibitem{Manohar:1992nd} 
  A.~V.~Manohar and M.~B.~Wise,
  %``Exotic Q Q anti-q anti-q states in QCD,''
  Nucl.\ Phys.\ B {\bf 399}, 17 (1993)
 % doi:10.1016/0550-3213(93)90614-U
  [hep-ph/9212236];
  %%CITATION = doi:10.1016/0550-3213(93)90614-U;%%
  %135 citations counted in INSPIRE as of 30 Jul 2016
 %\cite{Brambilla:2010cs}
%\bibitem{Brambilla:2010cs} 
  N.~Brambilla {\it et al.},
  %``Heavy quarkonium: progress, puzzles, and opportunities,''
  Eur.\ Phys.\ J.\ C {\bf 71}, 1534 (2011)
 % doi:10.1140/epjc/s10052-010-1534-9
  [arXiv:1010.5827 [hep-ph]];
  %%CITATION = doi:10.1140/epjc/s10052-010-1534-9;%%
  %986 citations counted in INSPIRE as of 19 Aug 2016
 %\bibitem{OTHERSXX}
  %%CITATION = ARXIV:1209.0144;%%
  %2 citations counted in INSPIRE as of 04 Aug 2016
%\cite{Voloshin:2007dx}
%\bibitem{Voloshin:2007dx} 
  M.~B.~Voloshin,
  %``Charmonium,''
  Prog.\ Part.\ Nucl.\ Phys.\  {\bf 61}, 455 (2008)
 % doi:10.1016/j.ppnp.2008.02.001
  [arXiv:0711.4556 [hep-ph]];
  %%CITATION = doi:10.1016/j.ppnp.2008.02.001;%%
  %276 citations counted in INSPIRE as of 19 Aug 2016
%\cite{Richard:2016eis}
%\bibitem{Richard:2016eis}
  J.~M.~Richard,
  %``Exotic hadrons: review and perspectives,''
  arXiv:1606.08593 [hep-ph].
  %%CITATION = ARXIV:1606.08593;%%


\bibitem{RISKA}
%\cite{Riska:1992qd}
%\bibitem{Riska:1992qd} 
  D.~O.~Riska and N.~N.~Scoccola,
  %``Anti-charm and anti-bottom hyperons,''
  Phys.\ Lett.\ B {\bf 299}, 338 (1993).
  %doi:10.1016/0370-2693(93)90270-R
  %%CITATION = doi:10.1016/0370-2693(93)90270-R;%%
  %78 citations counted in INSPIRE as of 30 Jul 2016

\bibitem{MACIEK2}
%\cite{Nowak:1993ff}
%\bibitem{Nowak:1993ff} 
  M.~A.~Nowak, I.~Zahed and M.~Rho,
  %``Heavy solitonic baryons,''
  Phys.\ Lett.\ B {\bf 303}, 130 (1993).
 % doi:10.1016/0370-2693(93)90056-N
  %%CITATION = doi:10.1016/0370-2693(93)90056-N;%%
  %20 citations counted in INSPIRE as of 30 Jul 2016

\bibitem{MACIEK3}
%\cite{Chernyshev:1995gj}
%\bibitem{Chernyshev:1995gj} 
  S.~Chernyshev, M.~A.~Nowak and I.~Zahed,
  %``Heavy hadrons and QCD instantons,''
  Phys.\ Rev.\ D {\bf 53}, 5176 (1996)
 % doi:10.1103/PhysRevD.53.5176
  [hep-ph/9510326].
  %%CITATION = doi:10.1103/PhysRevD.53.5176;%%
  %24 citations counted in INSPIRE as of 30 Jul 2016

\bibitem{SUHONG}
%\cite{Nielsen:2009uh}
%\bibitem{Nielsen:2009uh} 
  M.~Nielsen, F.~S.~Navarra and S.~H.~Lee,
  %``New Charmonium States in QCD Sum Rules: A Concise Review,''
  Phys.\ Rept.\  {\bf 497}, 41 (2010)
 % doi:10.1016/j.physrep.2010.07.005
  [arXiv:0911.1958 [hep-ph]].
  %%CITATION = doi:10.1016/j.physrep.2010.07.005;%%
  %175 citations counted in INSPIRE as of 19 Aug 2016

\bibitem{MAREK}
%\cite{Karliner:2015ina}
%\bibitem{Karliner:2015ina} 
  M.~Karliner and J.~L.~Rosner,
  %``New Exotic Meson and Baryon Resonances from Doubly-Heavy Hadronic Molecules,''
  Phys.\ Rev.\ Lett.\  {\bf 115}, no. 12, 122001 (2015)
%  doi:10.1103/PhysRevLett.115.122001
  [arXiv:1506.06386 [hep-ph]];
  %%CITATION = doi:10.1103/PhysRevLett.115.122001;%%
  %72 citations counted in INSPIRE as of 10 Apr 2017
%\cite{Karliner:2016joc}
%\bibitem{Karliner:2016joc} 
  M.~Karliner,
  %``Pentaquarks and doubly heavy exotic mesons,''
  EPJ Web Conf.\  {\bf 130}, 01003 (2016).
 % doi:10.1051/epjconf/201613001003
  %%CITATION = doi:10.1051/epjconf/201613001003;%%


\bibitem{MANY}
%\cite{Chen:2015loa}
%\bibitem{Chen:2015loa} 
  R.~Chen, X.~Liu, X.~Q.~Li and S.~L.~Zhu,
  %``Identifying exotic hidden-charm pentaquarks,''
  Phys.\ Rev.\ Lett.\  {\bf 115}, no. 13, 132002 (2015)
 % doi:10.1103/PhysRevLett.115.132002
  [arXiv:1507.03704 [hep-ph]];
  %%CITATION = doi:10.1103/PhysRevLett.115.132002;%%
  %84 citations counted in INSPIRE as of 10 Apr 2017
%\cite{Chen:2015moa}
%\bibitem{Chen:2015moa} 
  H.~X.~Chen, W.~Chen, X.~Liu, T.~G.~Steele and S.~L.~Zhu,
  %``Towards exotic hidden-charm pentaquarks in QCD,''
  Phys.\ Rev.\ Lett.\  {\bf 115}, no. 17, 172001 (2015)
 % doi:10.1103/PhysRevLett.115.172001
  [arXiv:1507.03717 [hep-ph]];
  %%CITATION = doi:10.1103/PhysRevLett.115.172001;%%
  %77 citations counted in INSPIRE as of 10 Apr 2017
%\cite{Roca:2015dva}
%\bibitem{Roca:2015dva} 
  L.~Roca, J.~Nieves and E.~Oset,
  %``LHCb pentaquark as a $\bar{D}^*\Sigma_c-\bar{D}^*\Sigma_c^*$ molecular state,''
  Phys.\ Rev.\ D {\bf 92}, no. 9, 094003 (2015)
 % doi:10.1103/PhysRevD.92.094003
  [arXiv:1507.04249 [hep-ph]];
  %%CITATION = doi:10.1103/PhysRevD.92.094003;%%
  %85 citations counted in INSPIRE as of 10 Apr 2017
  %\cite{Burns:2015dwa}
%\bibitem{Burns:2015dwa} 
  T.~J.~Burns,
  %``Phenomenology of P$_{c}$(4380)$^{+}$, P$_{c}$(4450)$^{+}$ and related states,''
  Eur.\ Phys.\ J.\ A {\bf 51}, no. 11, 152 (2015)
  %doi:10.1140/epja/i2015-15152-6
  [arXiv:1509.02460 [hep-ph]].
  %%CITATION = doi:10.1140/epja/i2015-15152-6;%%
  %47 citations counted in INSPIRE as of 10 Apr 2017
  %\cite{Huang:2015uda}
%\bibitem{Huang:2015uda} 
  H.~Huang, C.~Deng, J.~Ping and F.~Wang,
  %``Possible pentaquarks with heavy quarks,''
  Eur.\ Phys.\ J.\ C {\bf 76}, no. 11, 624 (2016)
  %doi:10.1140/epjc/s10052-016-4476-z
  [arXiv:1510.04648 [hep-ph]];
  %%CITATION = doi:10.1140/epjc/s10052-016-4476-z;%%
  %27 citations counted in INSPIRE as of 10 Apr 2017
 %\cite{Roca:2016tdh}
%\bibitem{Roca:2016tdh} 
  L.~Roca and E.~Oset,
  %``On the hidden charm pentaquarks in $\Lambda _b \rightarrow J/\psi K^- p$ decay,''
  Eur.\ Phys.\ J.\ C {\bf 76}, no. 11, 591 (2016)
  %doi:10.1140/epjc/s10052-016-4407-z
  [arXiv:1602.06791 [hep-ph]];
  %%CITATION = doi:10.1140/epjc/s10052-016-4407-z;%%
  %11 citations counted in INSPIRE as of 10 Apr 2017 
  %\cite{Lu:2016nnt}
%\bibitem{Lu:2016nnt} 
  Q.~F.~L and Y.~B.~Dong,
  %``Strong decay mode $J/\psi p$ of hidden charm pentaquark states $P_c^+(4380)$ and $P_c^+(4450)$ in $\Sigma_c \bar{D}^*$ molecular scenario,''
  Phys.\ Rev.\ D {\bf 93}, no. 7, 074020 (2016)
%  doi:10.1103/PhysRevD.93.074020
  [arXiv:1603.00559 [hep-ph]];
  %%CITATION = doi:10.1103/PhysRevD.93.074020;%%
  %17 citations counted in INSPIRE as of 10 Apr 2017
  %\cite{Shimizu:2016rrd}
%\bibitem{Shimizu:2016rrd} 
  Y.~Shimizu, D.~Suenaga and M.~Harada,
  %``Coupled channel analysis of molecule picture of $P_{c}(4380)$,''
  Phys.\ Rev.\ D {\bf 93}, no. 11, 114003 (2016)
  %doi:10.1103/PhysRevD.93.114003
  [arXiv:1603.02376 [hep-ph]];
  %%CITATION = doi:10.1103/PhysRevD.93.114003;%%
  %14 citations counted in INSPIRE as of 10 Apr 2017
  %\cite{Shen:2016tzq}
%\bibitem{Shen:2016tzq} 
  C.~W.~Shen, F.~K.~Guo, J.~J.~Xie and B.~S.~Zou,
  %``Disentangling the hadronic molecule nature of the $P_c(4380)$ pentaquark-like structure,''
  Nucl.\ Phys.\ A {\bf 954}, 393 (2016)
  %doi:10.1016/j.nuclphysa.2016.04.034
  [arXiv:1603.04672 [hep-ph]];
  %%CITATION = doi:10.1016/j.nuclphysa.2016.04.034;%%
  %12 citations counted in INSPIRE as of 10 Apr 2017
 %\cite{Eides:2015dtr}
%\bibitem{Eides:2015dtr} 
  M.~I.~Eides, V.~Y.~Petrov and M.~V.~Polyakov,
  %``Narrow nucleon-$\psi(2S)$ bound state and LHCb pentaquarks,''
  Phys.\ Rev.\ D {\bf 93}, no. 5, 054039 (2016)
 % doi:10.1103/PhysRevD.93.054039
  [arXiv:1512.00426 [hep-ph]];
  %%CITATION = doi:10.1103/PhysRevD.93.054039;%%
  %15 citations counted in INSPIRE as of 10 Apr 2017
 %\cite{Perevalova:2016dln}
%\bibitem{Perevalova:2016dln} 
  I.~A.~Perevalova, M.~V.~Polyakov and P.~Schweitzer,
  %``On LHCb pentaquarks as a baryon-$\psi$(2S) bound state: prediction of isospin-$\frac3{2}$ pentaquarks with hidden charm,''
  Phys.\ Rev.\ D {\bf 94}, no. 5, 054024 (2016)
  %doi:10.1103/PhysRevD.94.054024
  [arXiv:1607.07008 [hep-ph]];
  %%CITATION = doi:10.1103/PhysRevD.94.054024;%%
  %2 citations counted in INSPIRE as of 10 Apr 2017
  %\cite{Kopeliovich:2016gra}
%\bibitem{Kopeliovich:2016gra} 
  V.~Kopeliovich and I.~Potashnikova,
  %``Simple estimates of the masses of pentaquarks with hidden beauty or strangeness,''
  Phys.\ Rev.\ D {\bf 93}, no. 7, 074012 (2016);
 % doi:10.1103/PhysRevD.93.074012
  %%CITATION = doi:10.1103/PhysRevD.93.074012;%%
  %3 citations counted in INSPIRE as of 10 Apr 2017
  %\cite{Yamaguchi:2016ote}
%\bibitem{Yamaguchi:2016ote} 
  Y.~Yamaguchi and E.~Santopinto,
  %``Hidden-charm pentaquarks as a meson-baryon molecule with coupled channels for $\bar{D}^{(\ast)}\Lambda_{\rm c}$ and $\bar{D}^{(\ast)}\Sigma^{(\ast)}_{\rm c}$,''
  arXiv:1606.08330 [hep-ph];
  %%CITATION = ARXIV:1606.08330;%%
  %9 citations counted in INSPIRE as of 10 Apr 2017
  %\cite{Takeuchi:2016ejt}
%\bibitem{Takeuchi:2016ejt} 
  S.~Takeuchi and M.~Takizawa,
  %``The hidden charm pentaquarks are the hidden color-octet $uud$ baryons?,''
  Phys.\ Lett.\ B {\bf 764}, 254 (2017)
 % doi:10.1016/j.physletb.2016.11.034
  [arXiv:1608.05475 [hep-ph]].
  %%CITATION = doi:10.1016/j.physletb.2016.11.034;%%
  %3 citations counted in INSPIRE as of 10 Apr 2017
  
  
  \bibitem{PENTARHO}
  %\cite{Scoccola:2015nia}
%\bibitem{Scoccola:2015nia} 
  N.~N.~Scoccola, D.~O.~Riska and M.~Rho,
  %``Pentaquark candidates P$_c^+$(4380) and P$_c^+$(4450) within the soliton picture of baryons,''
  Phys.\ Rev.\ D {\bf 92}, no. 5, 051501 (2015)
 % doi:10.1103/PhysRevD.92.051501
  [arXiv:1508.01172 [hep-ph]].
  %%CITATION = doi:10.1103/PhysRevD.92.051501;%%
  %40 citations counted in INSPIRE as of 10 Apr 2017
  
  

 
 \bibitem{VENEZIANO}
  %\cite{Rossi:2016szw}
%\bibitem{Rossi:2016szw} 
  G.~Rossi and G.~Veneziano,
  %``The string-junction picture of multiquark states: an update,''
  JHEP {\bf 1606}, 041 (2016)
  %doi:10.1007/JHEP06(2016)041
  [arXiv:1603.05830 [hep-th]].
  %%CITATION = doi:10.1007/JHEP06(2016)041;%%
  %13 citations counted in INSPIRE as of 10 Apr 2017


\bibitem{COBI}
%\cite{Sonnenschein:2016ibx}
%\bibitem{Sonnenschein:2016ibx} 
  J.~Sonnenschein and D.~Weissman,
  %``A tetraquark or not a tetraquark: A holography inspired stringy hadron (HISH) perspective,''
  arXiv:1606.02732 [hep-ph].
  %%CITATION = ARXIV:1606.02732;%%
  %5 citations counted in INSPIRE as of 10 Apr 2017








\bibitem{SSX}
%\cite{Sakai:2004cn}
%\bibitem{Sakai:2004cn} 
  T.~Sakai and S.~Sugimoto,
  %``Low energy hadron physics in holographic QCD,''
  Prog.\ Theor.\ Phys.\  {\bf 113}, 843 (2005)
 % doi:10.1143/PTP.113.843
  [hep-th/0412141];
  %%CITATION = doi:10.1143/PTP.113.843;%%
  %1001 citations counted in INSPIRE as of 01 Sep 2016
%\cite{Sakai:2005yt}
%\bibitem{Sakai:2005yt} 
  T.~Sakai and S.~Sugimoto,
  %``More on a holographic dual of QCD,''
  Prog.\ Theor.\ Phys.\  {\bf 114}, 1083 (2005)
%  doi:10.1143/PTP.114.1083
  [hep-th/0507073].
  %%CITATION = doi:10.1143/PTP.114.1083;%%
  %574 citations counted in INSPIRE as of 01 Sep 2016


\bibitem{HIDDEN}
%\cite{Fujiwara:1984mp}
%\bibitem{Fujiwara:1984mp} 
  T.~Fujiwara, T.~Kugo, H.~Terao, S.~Uehara and K.~Yamawaki,
  %``Nonabelian Anomaly and Vector Mesons as Dynamical Gauge Bosons of Hidden Local Symmetries,''
  Prog.\ Theor.\ Phys.\  {\bf 73}, 926 (1985).
  %doi:10.1143/PTP.73.926
  %%CITATION = doi:10.1143/PTP.73.926;%%
  %221 citations counted in INSPIRE as of 07 Sep 2016



\bibitem{FEWX}
%\cite{Paredes:2004is}
%\bibitem{Paredes:2004is} 
  A.~Paredes and P.~Talavera,
  %``Multiflavor excited mesons from the fifth dimension,''
  Nucl.\ Phys.\ B {\bf 713}, 438 (2005)
  %doi:10.1016/j.nuclphysb.2005.02.021
  [hep-th/0412260];
  %%CITATION = doi:10.1016/j.nuclphysb.2005.02.021;%%
  %41 citations counted in INSPIRE as of 14 Nov 2016
%\cite{Erdmenger:2006bg}
%\bibitem{Erdmenger:2006bg} 
  J.~Erdmenger, N.~Evans and J.~Grosse,
  %``Heavy-light mesons from the AdS/CFT correspondence,''
  JHEP {\bf 0701}, 098 (2007);
  %doi:10.1088/1126-6708/2007/01/098
  [hep-th/0605241].
  %%CITATION = doi:10.1088/1126-6708/2007/01/098;%%
  %29 citations counted in INSPIRE as of 01 Sep 2016
  %\cite{Erdmenger:2007vj}
%\bibitem{Erdmenger:2007vj}
  J.~Erdmenger, K.~Ghoroku and I.~Kirsch,
  %``Holographic heavy-light mesons from non-Abelian DBI,''
  JHEP {\bf 0709} (2007) 111
 % doi:10.1088/1126-6708/2007/09/111
  [arXiv:0706.3978 [hep-th]];
  %%CITATION = doi:10.1088/1126-6708/2007/09/111;%%
  %23 citations counted in INSPIRE as of 01 Sep 2016
 %\cite{Herzog:2008bp}
%\bibitem{Herzog:2008bp} 
  C.~P.~Herzog, S.~A.~Stricker and A.~Vuorinen,
  %``Remarks on Heavy-Light Mesons from AdS/CFT,''
  JHEP {\bf 0805}, 070 (2008)
 % doi:10.1088/1126-6708/2008/05/070
  [arXiv:0802.2956 [hep-th]];
  %%CITATION = doi:10.1088/1126-6708/2008/05/070;%%
  %9 citations counted in INSPIRE as of 01 Sep 2016
  %\cite{Bai:2013rza}
%\bibitem{Bai:2013rza} 
  Y.~Bai and H.~C.~Cheng,
  %``A Holographic Model of Heavy-light Mesons,''
  JHEP {\bf 1308}, 074 (2013)
 % doi:10.1007/JHEP08(2013)074
  [arXiv:1306.2944 [hep-ph]];
  %%CITATION = doi:10.1007/JHEP08(2013)074;%%
  %1 citations counted in INSPIRE as of 01 Sep 2016
  %\cite{Hashimoto:2014jua}
%\bibitem{Hashimoto:2014jua} 
  K.~Hashimoto, N.~Ogawa and Y.~Yamaguchi,
  %``Holographic Heavy Quark Symmetry,''
  JHEP {\bf 1506}, 040 (2015)
  %doi:10.1007/JHEP06(2015)040
  [arXiv:1412.5590 [hep-th]].
  %%CITATION = doi:10.1007/JHEP06(2015)040;%%
  %3 citations counted in INSPIRE as of 01 Sep 2016
%\cite{Sonnenschein:2016ibx}
%\bibitem{Sonnenschein:2016ibx} 
  J.~Sonnenschein and D.~Weissman,
  %``A tetraquark or not a tetraquark: A holography inspired stringy hadron (HISH) perspective,''
  arXiv:1606.02732 [hep-ph].
  %%CITATION = ARXIV:1606.02732;%%
  %1 citations counted in INSPIRE as of 01 Sep 2016


\bibitem{BRODSKY}
%\cite{deTeramond:2016htp}
%5\bibitem{deTeramond:2016htp} 
  G.~F.~de Teramond, S.~J.~Brodsky, A.~Deur, H.~G.~Dosch and R.~S.~Sufian,
  %``Superconformal Algebraic Approach to Hadron Structure,''
  arXiv:1611.03763 [hep-ph];
  %%CITATION = ARXIV:1611.03763;%%
%\cite{Dosch:2015bca}
%\bibitem{Dosch:2015bca}
  H.~G.~Dosch, G.~F.~de Teramond and S.~J.~Brodsky,
  %``Supersymmetry Across the Light and Heavy-Light Hadronic Spectrum,''
  Phys.\ Rev.\ D {\bf 92} (2015) no.7,  074010
 % doi:10.1103/PhysRevD.92.074010
  [arXiv:1504.05112 [hep-ph]];
  %%CITATION = doi:10.1103/PhysRevD.92.074010;%%
  %17 citations counted in INSPIRE as of 17 Nov 2016
H.~G.~Dosch, G.~F.~de Teramond and S. J. Brodsky,
  %``Supersymmetry Across the Light and Heavy-Light Hadronic Spectrum II,
  Phys.\ Rev. \ D {\bf 95} (2017) no. 3, 034016
  [arXiv:1612.02370 [hep-ph]].





\bibitem{MEYERS}
%\cite{Myers:1999ps}
%\bibitem{Myers:1999ps} 
  R.~C.~Myers,
  %``Dielectric branes,''
  JHEP {\bf 9912}, 022 (1999)
%  doi:10.1088/1126-6708/1999/12/022
  [hep-th/9910053].
  %%CITATION = doi:10.1088/1126-6708/1999/12/022;%%
  %1143 citations counted in INSPIRE as of 01 Sep 2016


\bibitem{SSXB}
%\cite{Hata:2007mb}
%\bibitem{Hata:2007mb}
  H.~Hata, T.~Sakai, S.~Sugimoto and S.~Yamato,
  %``Baryons from instantons in holographic QCD,''
  Prog.\ Theor.\ Phys.\  {\bf 117} (2007) 1157
 %doi:10.1143/PTP.117.1157
  [hep-th/0701280 [HEP-TH]].
  %%CITATION = doi:10.1143/PTP.117.1157;%%
  %215 citations counted in INSPIRE as of 05 Apr 2017


\bibitem{SSXBB}
%\cite{Hashimoto:2008zw}
%\bibitem{Hashimoto:2008zw}
  K.~Hashimoto, T.~Sakai and S.~Sugimoto,
  %``Holographic Baryons: Static Properties and Form Factors from Gauge/String Duality,''
  Prog.\ Theor.\ Phys.\  {\bf 120} (2008) 1093
  %doi:10.1143/PTP.120.1093
  [arXiv:0806.3122 [hep-th]];
  %%CITATION = doi:10.1143/PTP.120.1093;%%
  %128 citations counted in INSPIRE as of 05 Apr 2017
%\cite{Kim:2008pw}
%\bibitem{Kim:2008pw} 
  K.~Y.~Kim and I.~Zahed,
  %``Electromagnetic Baryon Form Factors from Holographic QCD,''
  JHEP {\bf 0809}, 007 (2008)
 % doi:10.1088/1126-6708/2008/09/007
  [arXiv:0807.0033 [hep-th]].
  %%CITATION = doi:10.1088/1126-6708/2008/09/007;%%
  %35 citations counted in INSPIRE as of 05 Apr 2017

%\bibitem{LIU-BUP}
%Y.~Liu and I.~Zahed, in preparation.
  
  
  
  \bibitem{SKYRME}
  %\cite{Zahed:1986qz}
%\bibitem{Zahed:1986qz} 
  I.~Zahed and G.~E.~Brown,
  %``The Skyrme Model,''
  Phys.\ Rept.\  {\bf 142}, 1 (1986);
%  doi:10.1016/0370-1573(86)90142-0
  %%CITATION = doi:10.1016/0370-1573(86)90142-0;%%
  %577 citations counted in INSPIRE as of 10 Apr 2017
   Multifaceted Skyrmion, Eds. M.~Rho and I.~Zahed, World Scientific, 2016.
  
 \bibitem{LIUHEAVY}
 %\cite{Liu:2016iqo}
%\bibitem{Liu:2016iqo} 
  Y.~Liu and I.~Zahed,
  %``Holographic Heavy-Light Chiral Effective Action,''
  Phys.\ Rev.\ D {\bf 95}, no. 5, 056022 (2017)
%  doi:10.1103/PhysRevD.95.056022
  [arXiv:1611.03757 [hep-ph]].
  %%CITATION = doi:10.1103/PhysRevD.95.056022;%%
  %3 citations counted in INSPIRE as of 10 Apr 2017
 %\cite{Liu:2016urz}
%\bibitem{Liu:2016urz} 
  Y.~Liu and I.~Zahed,
  %``Heavy-Light Mesons in Chiral AdS/QCD,''
  arXiv:1611.04400 [hep-ph].
  %%CITATION = ARXIV:1611.04400;%%
  %3 citations counted in INSPIRE as of 10 Apr 2017




 \bibitem{SKYRMEHEAVY} 
 %\cite{Scoccola:1991te}
%\bibitem{Scoccola:1991te} 
  N.~N.~Scoccola,
  %``Heavy baryons in a topological soliton model,''
  Nucl.\ Phys.\ A {\bf 532}, 409C (1991);
%  doi:10.1016/0375-9474(91)90717-K
  %%CITATION = doi:10.1016/0375-9474(91)90717-K;%%
 %\cite{Schat:1999dw}
 %\cite{Rho:1992yy}
%\bibitem{Rho:1992yy} 
  M.~Rho, D.~O.~Riska and N.~N.~Scoccola,
  %``The Energy levels of the heavy flavor baryons in the topological soliton model,''
  Z.\ Phys.\ A {\bf 341}, 343 (1992);
 % doi:10.1007/BF01283544
  %%CITATION = doi:10.1007/BF01283544;%%
  %85 citations counted in INSPIRE as of 10 Apr 2017
 %\cite{Min:1992uk}
%\bibitem{Min:1992uk} 
  D.~P.~Min, Y.~s.~Oh, B.~Y.~Park and M.~Rho,
  %``Soliton structure of heavy baryons,''
  hep-ph/9209275.
  %%CITATION = HEP-PH/9209275;%%
  %17 citations counted in INSPIRE as of 10 Apr 2017
%\cite{Oh:1994vd}
%\bibitem{Oh:1994vd} 
  Y.~s.~Oh, B.~Y.~Park and D.~P.~Min,
  %``Heavy baryons as Skyrmion with 1/m(Q) corrections,''
  Phys.\ Rev.\ D {\bf 49}, 4649 (1994)
 % doi:10.1103/PhysRevD.49.4649
  [hep-ph/9402205];
  %%CITATION = doi:10.1103/PhysRevD.49.4649;%%
  %24 citations counted in INSPIRE as of 10 Apr 2017
 %\cite{Oh:1994yv}
%\bibitem{Oh:1994yv} 
  Y.~s.~Oh, B.~Y.~Park and D.~P.~Min,
  %``Heavy quark symmetry and the Skyrme model,''
  Phys.\ Rev.\ D {\bf 50}, 3350 (1994)
  %doi:10.1103/PhysRevD.50.3350
  [hep-ph/9407214];
  %%CITATION = doi:10.1103/PhysRevD.50.3350;%%
  %64 citations counted in INSPIRE as of 10 Apr 2017
%\cite{Min:1994qq}
%\bibitem{Min:1994qq} 
  D.~P.~Min, Y.~s.~Oh, B.~Y.~Park and M.~Rho,
  %``Heavy quark symmetry and skyrmions,''
  Int.\ J.\ Mod.\ Phys.\ E {\bf 4}, 47 (1995)
  %doi:10.1142/S0218301395000031
  [hep-ph/9412302];
  %%CITATION = doi:10.1142/S0218301395000031;%%
  %35 citations counted in INSPIRE as of 10 Apr 2017
%\cite{Oh:1994ux}
%\bibitem{Oh:1994ux} 
  Y.~s.~Oh and B.~Y.~Park,
  %``Energy levels of the soliton - heavy meson bound states,''
  Phys.\ Rev.\ D {\bf 51}, 5016 (1995)
 % doi:10.1103/PhysRevD.51.5016
  [hep-ph/9501356];
  %%CITATION = doi:10.1103/PhysRevD.51.5016;%%
  %29 citations counted in INSPIRE as of 10 Apr 2017
%\cite{Oh:1997tp}
%\cite{Schechter:1995vr}
%\bibitem{Schechter:1995vr} 
  J.~Schechter, A.~Subbaraman, S.~Vaidya and H.~Weigel,
  %``Heavy quark solitons: Towards realistic masses,''
  Nucl.\ Phys.\ A {\bf 590}, 655 (1995)
  Erratum: [Nucl.\ Phys.\ A {\bf 598}, 583 (1996)]
 % doi:10.1016/0375-9474(95)00182-Z, 10.1016/0375-9474(96)00013-9
  [hep-ph/9503307];
  %%CITATION = doi:10.1016/0375-9474(95)00182-Z, 10.1016/0375-9474(96)00013-9;%%
  %27 citations counted in INSPIRE as of 10 Apr 2017
%\bibitem{Oh:1997tp} 
  Y.~s.~Oh and B.~Y.~Park,
  %``Solitons bound to heavy mesons,''
  Z.\ Phys.\ A {\bf 359}, 83 (1997)
  %doi:10.1007/s002180050370
  [hep-ph/9703219];
  %%CITATION = doi:10.1007/s002180050370;%%
  %21 citations counted in INSPIRE as of 10 Apr 2017
%\bibitem{Schat:1999dw} 
  C.~L.~Schat and N.~N.~Scoccola,
  %``Multibaryons with heavy flavors in the Skyrme model,''
  Phys.\ Rev.\ D {\bf 61}, 034008 (2000)
 % doi:10.1103/PhysRevD.61.034008
  [hep-ph/9907271];
  %%CITATION = doi:10.1103/PhysRevD.61.034008;%%
  %9 citations counted in INSPIRE as of 10 Apr 2017
  %\cite{Scoccola:2009au}
%\bibitem{Scoccola:2009au} 
  N.~N.~Scoccola,
  %``Heavy quark skyrmions,''
  %doi:10.1142/9789814280709_0004
  arXiv:0905.2722 [hep-ph];
  %%CITATION = doi:10.1142/9789814280709_0004;%%
  %3 citations counted in INSPIRE as of 10 Apr 2017
 %\cite{Blanckenberg:2015dsa}
%\bibitem{Blanckenberg:2015dsa} 
  J.~P.~Blanckenberg and H.~Weigel,
  %``Heavy Baryons with Strangeness in a Soliton Model,''
  Phys.\ Lett.\ B {\bf 750}, 230 (2015)
  %doi:10.1016/j.physletb.2015.09.026
  [arXiv:1505.06655 [hep-ph]].
  %%CITATION = doi:10.1016/j.physletb.2015.09.026;%%
  %4 citations counted in INSPIRE as of 10 Apr 2017


\bibitem{THETA}
%\cite{Itzhaki:2003nr}
%\bibitem{Itzhaki:2003nr} 
  N.~Itzhaki, I.~R.~Klebanov, P.~Ouyang and L.~Rastelli,
  %``Is Theta+(1540) a kaon Skyrmion resonance?,''
  Nucl.\ Phys.\ B {\bf 684}, 264 (2004)
 % doi:10.1016/j.nuclphysb.2004.02.004
  [hep-ph/0309305].
  %%CITATION = doi:10.1016/j.nuclphysb.2004.02.004;%%
  %117 citations counted in INSPIRE as of 10 Apr 2017



\bibitem{HOLOXX}
%\cite{Maldacena:1997re}
%\bibitem{Maldacena:1997re} 
  J.~M.~Maldacena,
  %``The Large N limit of superconformal field theories and supergravity,''
  Int.\ J.\ Theor.\ Phys.\  {\bf 38}, 1113 (1999)
  [Adv.\ Theor.\ Math.\ Phys.\  {\bf 2}, 231 (1998)]
  %doi:10.1023/A:1026654312961
  [hep-th/9711200];
  %%CITATION = doi:10.1023/A:1026654312961;%%
  %12143 citations counted in INSPIRE as of 01 Oct 2016
%\cite{Gubser:1998bc}
%\bibitem{Gubser:1998bc} 
  S.~S.~Gubser, I.~R.~Klebanov and A.~M.~Polyakov,
  %``Gauge theory correlators from noncritical string theory,''
  Phys.\ Lett.\ B {\bf 428}, 105 (1998)
 % doi:10.1016/S0370-2693(98)00377-3
  [hep-th/9802109];
  %%CITATION = doi:10.1016/S0370-2693(98)00377-3;%%
  %6912 citations counted in INSPIRE as of 01 Oct 2016
%\cite{Witten:1998zw}
%\bibitem{Witten:1998zw} 
  E.~Witten,
  %``Anti-de Sitter space, thermal phase transition, and confinement in gauge theories,''
  Adv.\ Theor.\ Math.\ Phys.\  {\bf 2}, 505 (1998)
  [hep-th/9803131];
  %%CITATION = HEP-TH/9803131;%%
  %2432 citations counted in INSPIRE as of 01 Oct 2016
%\cite{Klebanov:1999tb}
%\bibitem{Klebanov:1999tb} 
  I.~R.~Klebanov and E.~Witten,
  %``AdS / CFT correspondence and symmetry breaking,''
  Nucl.\ Phys.\ B {\bf 556}, 89 (1999)
 % doi:10.1016/S0550-3213(99)00387-9
  [hep-th/9905104].
  %%CITATION = doi:10.1016/S0550-3213(99)00387-9;%%
  %835 citations counted in INSPIRE as of 31 Oct 2016



\bibitem{HOLOXXX}
%\cite{Erlich:2005qh}
%\bibitem{Erlich:2005qh} 
  J.~Erlich, E.~Katz, D.~T.~Son and M.~A.~Stephanov,
  %``QCD and a holographic model of hadrons,''
  Phys.\ Rev.\ Lett.\  {\bf 95}, 261602 (2005)
 % doi:10.1103/PhysRevLett.95.261602
  [hep-ph/0501128];
  %%CITATION = doi:10.1103/PhysRevLett.95.261602;%%
  %767 citations counted in INSPIRE as of 26 Oct 2016
%\cite{DaRold:2005mxj}
%\bibitem{DaRold:2005mxj} 
  L.~Da Rold and A.~Pomarol,
  %``Chiral symmetry breaking from five dimensional spaces,''
  Nucl.\ Phys.\ B {\bf 721}, 79 (2005)
  %doi:10.1016/j.nuclphysb.2005.05.009
  [hep-ph/0501218].
  %%CITATION = doi:10.1016/j.nuclphysb.2005.05.009;%%
  %565 citations counted in INSPIRE as of 26 Oct 2016





\bibitem{HOLOXXXX}
%\cite{Hong:2004sa}
%\bibitem{Hong:2004sa} 
  S.~Hong, S.~Yoon and M.~J.~Strassler,
  %``On the couplings of vector mesons in AdS / QCD,''
  JHEP {\bf 0604}, 003 (2006)
 % doi:10.1088/1126-6708/2006/04/003
  [hep-th/0409118];
  %%CITATION = doi:10.1088/1126-6708/2006/04/003;%%
  %86 citations counted in INSPIRE as of 01 Oct 2016
%\cite{Erlich:2006hq}
%\bibitem{Erlich:2006hq} 
  J.~Erlich, G.~D.~Kribs and I.~Low,
  %``Emerging holography,''
  Phys.\ Rev.\ D {\bf 73}, 096001 (2006)
  doi:10.1103/PhysRevD.73.096001
  [hep-th/0602110];
  %%CITATION = doi:10.1103/PhysRevD.73.096001;%%
  %48 citations counted in INSPIRE as of 01 Oct 2016
%\cite{Grigoryan:2007my}
%\bibitem{Grigoryan:2007my} 
  H.~R.~Grigoryan and A.~V.~Radyushkin,
  %``Structure of vector mesons in holographic model with linear confinement,''
  Phys.\ Rev.\ D {\bf 76}, 095007 (2007)
 % doi:10.1103/PhysRevD.76.095007
  [arXiv:0706.1543 [hep-ph]];
  %%CITATION = doi:10.1103/PhysRevD.76.095007;%%
  %127 citations counted in INSPIRE as of 01 Oct 2016
%\cite{Grigoryan:2007vg}
%\bibitem{Grigoryan:2007vg} 
  H.~R.~Grigoryan and A.~V.~Radyushkin,
  %``Form Factors and Wave Functions of Vector Mesons in Holographic QCD,''
  Phys.\ Lett.\ B {\bf 650}, 421 (2007)
 % doi:10.1016/j.physletb.2007.05.044
  [hep-ph/0703069];
  %%CITATION = doi:10.1016/j.physletb.2007.05.044;%%
  %100 citations counted in INSPIRE as of 09 Nov 2016
%\cite{Afonin:2016lij}
%\bibitem{Afonin:2016lij} 
  S.~S.~Afonin and I.~V.~Pusenkov,
  %``The no-wall holographic model for vector quarkonia,''
  EPJ Web Conf.\  {\bf 125}, 04004 (2016)
  %doi:10.1051/epjconf/201612504004
  [arXiv:1606.06091 [hep-ph]];
  %%CITATION = doi:10.1051/epjconf/201612504004;%%
%\cite{Braga:2015lck}
%\bibitem{Braga:2015lck} 
  N.~R.~F.~Braga, M.~A.~Martin Contreras and S.~Diles,
  %``Holographic model for heavy-vector-meson masses,''
  Europhys.\ Lett.\  {\bf 115}, no. 3, 31002 (2016)
%  doi:10.1209/0295-5075/115/31002
  [arXiv:1511.06373 [hep-th]];
  %%CITATION = doi:10.1209/0295-5075/115/31002;%%
  %1 citations counted in INSPIRE as of 10 Nov 2016
%\cite{Gorsky:2015pra}
%\bibitem{Gorsky:2015pra} 
  A.~Gorsky, S.~B.~Gudnason and A.~Krikun,
  %``Baryon and chiral symmetry breaking in holographic QCD,''
  Phys.\ Rev.\ D {\bf 91}, no. 12, 126008 (2015)
%  doi:10.1103/PhysRevD.91.126008
  [arXiv:1503.04820 [hep-th]].
  %%CITATION = doi:10.1103/PhysRevD.91.126008;%%
  %1 citations counted in INSPIRE as of 10 Nov 2016




%\bibitem{PDG}
%Particle data group, pdg.lbl.gov/








%\bibitem{AOKI}
%A. Aoki et al., Phys. Rev. Lett. {\bf 80}, 5711 (1998).


\bibitem{KIRITSIS}
%\cite{Gursoy:2007cb}
%\bibitem{Gursoy:2007cb} 
  U.~Gursoy and E.~Kiritsis,
  %``Exploring improved holographic theories for QCD: Part I,''
  JHEP {\bf 0802}, 032 (2008)
 % doi:10.1088/1126-6708/2008/02/032
  [arXiv:0707.1324 [hep-th]];
  %%CITATION = doi:10.1088/1126-6708/2008/02/032;%%
  %260 citations counted in INSPIRE as of 06 Oct 2016
%\cite{Gursoy:2007er}
%\bibitem{Gursoy:2007er} 
  U.~Gursoy, E.~Kiritsis and F.~Nitti,
  %``Exploring improved holographic theories for QCD: Part II,''
  JHEP {\bf 0802}, 019 (2008)
 % doi:10.1088/1126-6708/2008/02/019
  [arXiv:0707.1349 [hep-th]].
  %%CITATION = doi:10.1088/1126-6708/2008/02/019;%%
  %270 citations counted in INSPIRE as of 06 Oct 2016






\bibitem{ADAMI}
%\cite{Adami:1987xz}
%\bibitem{Adami:1987xz} 
  C.~Adami and I.~Zahed,
  %``The Width of the $\Delta$ Isobar in Chiral Soliton Models,''
  Phys.\ Lett.\ B {\bf 213}, 373 (1988).
 % doi:10.1016/0370-2693(88)91778-9
  %%CITATION = doi:10.1016/0370-2693(88)91778-9;%%
  %18 citations counted in INSPIRE as of 11 Apr 2017















\end{thebibliography}
\end{document}